\documentclass[fleqn,usenatbib]{mnras}

\usepackage{newtxtext,newtxmath}

\usepackage[T1]{fontenc}
\usepackage{ae,aecompl,amsmath}
\usepackage{xspace}
\usepackage[dvipsnames]{xcolor}

\usepackage{graphicx}	
\usepackage{amsmath}	
\usepackage{comment}    
\usepackage{float}
\usepackage{epstopdf}
\usepackage[colorinlistoftodos]{todonotes}  %
\usepackage{orcidlink}
\newcommand{\angstrom}{\textup{\AA}}

\title[Metallicity from UV and optical lines-II]{Metallicity of Active Galactic Nuclei from ultraviolet and optical emission lines
– II.  Revisiting the $C43$ metallicity calibration and its implications}
\author[O.~L. Dors et al.]{O.~L. Dors$^{1}$\thanks{E-mail: olidors@univap.br}   \orcidlink{0000-0003-4782-1570}, C.~B. Oliveira$^{1}$, M.~V. Cardaci$^{2,3}$, G.~F. H\"agele$^{2,3}$, 
Mark Armah$^{1}$, R.~A. Riffel$^{4}$, \newauthor{L.~Ramos Vieira$^{4,5}$, G.~C. Almeida$^{1}$, I.~N. Morais$^{1}$, P.~C. Santos$^{1}$}
\\
$^{1}$Universidade do Vale do Para\'iba, Av. Shishima Hifumi, 2911, Cep
12244-000, S\~ao Jos\'e dos Campos, SP, Brazil \\
$^2$ Facultad de Ciencias Astron\'omicas y Geof\'{\i}sicas, Universidad Nacional de La Plata, Paseo del Bosque s/n, 1900 La Plata, Argentina.\\
$^3$ Instituto de Astrofísica de La Plata (CONICET-UNLP), La Plata, Avenida Centenario (Paseo del Bosque) S/N, B1900FWA, Argentina\\
$^{4}$ Departamento de F\'isica, Centro de Ci\^encias Naturais e Exatas, Universidade Federal de Santa Maria, 97105-900, Santa Maria, RS, Brazil\\
$^{5}$ Instituto Federal Catarinense - Campus Concórdia, 89703-720, Concórdia, SC, Brazil\\
}

\date{Accepted XXX. Received YYY; in original form ZZZ}

\pubyear{2025}

\begin{document}
\label{firstpage}
\pagerange{\pageref{firstpage}--\pageref{lastpage}}
\maketitle

\begin{abstract}
In this study, a new semi-empirical calibration is proposed between ultraviolet emission lines (\ion{C}{iii}]$\lambda1909$, \ion{C}{iv}$\lambda1549$, \ion{He}{ii}]$\lambda1640$) of type~2 AGNs and their metallicity ($Z$). This calibration is derived by comparing a large sample of 106 objects (data taken from the literature) located over a wide range of redshifts ($0 \: \lesssim \: z \: \lesssim \: 4.0$) with predictions from photoionization models that adopt a recent C/O–O/H relation derived via estimates using the $T_{\rm e}$ method, which is considered the most reliable method. We found that the new calibration produces $Z$ values in agreement (within an uncertainty of $\pm 0.1$ dex) with those from other calibrations and from estimates via the $T_{\rm e}$-method.   We find also that AGN metallicities are already high at early epochs, with no evidence for monotonic evolution across the redshift range $0 \: \lesssim \: z \: \lesssim \: 12$. Notably, the highest metallicities in our sample, reaching up to $\rm 4\: Z_{\odot}$, are found in objects at $2 \lesssim z \lesssim 3$. This redshift range coincides with the peak of the cosmic star formation rate history, suggesting a strong connection between the major epoch of star formation, black hole growth, and rapid metal enrichment in the host galaxies of  AGNs. Furthermore,  our analysis reveals no significant correlation between AGN metallicity and radio properties (radio spectral index or radio luminosity) or host galaxy stellar mass. The lack of a clear mass-metallicity relation, consistent with findings for local AGNs, suggests that the chemical evolution of the nuclear gas is decoupled from the global properties of the host galaxy.
\end{abstract}

\begin{keywords}
galaxies: Seyfert -- galaxies: active -- galaxies: abundances --ISM: abundances
--galaxies: evolution --galaxies: nuclei 
\end{keywords}



\section{Introduction}

Metallicity ($Z$) estimations based on emission lines are essential for understanding the chemical evolution of galaxies, stellar nucleosynthesis and the interplay between stars and the interstellar medium (ISM) over the Hubble time. Observations from ground-based telescopes, combined with recent data from the James Webb Space Telescope (JWST), have enabled the estimation of $Z$ in the gas-phase of objects across a wide range of redshifts \citep[e.g. $0 \: \leq z \: \lesssim \: 14$;][]{2024Natur.633..318C}, re\-vo\-lu\-tionizing our understanding of the chemical enrichment of the Universe.

The methods for estimating $Z$ depend on the set of emission lines available and on their
wavelength range (for a review, see \citealt{2019A&ARv..27....3M, 2019ARA&A..57..511K}).
For objects at $z \: \lesssim \: 0.4$, where most observational data are in the optical range [$3000 \: < \: \lambda(\angstrom) \:  < \: 7000$; e.g. \citealt{2000AJ....120.1579Y, 2012A&A...538A...8S, 2015ApJ...798....7B, 2015ApJ...806...16B}], the most reliable method is the $T_{\rm e}$-method or the direct method, known to have been developed by \citet{1969BOTT....5....3P}. 
Summarily, this method is based on the estimation of
the electron temperature through auroral lines
(e.g. [\ion{O}{iii}]$\lambda4363$), which are about 100 times
weaker than H$\beta$ and measured mainly in
spectra of objects with low metallicity ($Z \: \lesssim \: \rm Z_{\odot}$) and/or with high excitation 
(e.g. \citealt{1998AJ....116.2805V, 2003ApJ...591..801K, 2008MNRAS.383..209H, 2022ApJ...939...44R, 2025A&A...694A..61G}).
To circumvent the difficulty of applying the $T_{\rm e}$-method,
\citet{1979MNRAS.189...95P}, following the original idea
of \citet{1976ApJ...209..748J}, proposed the strong-line method --- a calibration between
the $R_{23}$=([\ion{O}{ii}]$\lambda3727$+[\ion{O}{ii}]$\lambda4959+\lambda5007$)/H$\beta$ line ratio
and the oxygen abundance in relation
to hydrogen (O/H, a $Z$ tracer). 
Both the $T_{\rm e}$-method and the strong-line methods have been
adapted for active galactic nuclei (AGNs) by  \citet{2020MNRAS.496.3209D} and
\citet{1998AJ....115..909S}, respectively. 
In particular, AGNs are bright objects that can be used
to trace the $Z$ evolution of the Universe instead of
star-forming galaxies, hence a large number of active
galaxies has been observed in the local
universe (e.g. \citealt{1978ApJ...223...56K, 2015ApJS..217...12D, 2018A&A...618A...6K, 2018ApJ...856...46R, 2019MNRAS.489.3351B, 2021MNRAS.503.2583S, 2022MNRAS.509.2377D, 2022MNRAS.513.5935M}) and
at $z \: \gtrsim \: 5$ (e.g. \citealt{2023MNRAS.525.1353J, 2023arXiv231118731S, 2024A&A...691A.145M, 2024arXiv241104944D, 2025A&A...693A..50N, 2025arXiv250303431D, 2025arXiv250622147G}).

Calibrations between emission line intensity ratios and $Z$ 
can be obtained, mainly, by three methods, each with its own limitations, as summarized below.
\begin{itemize}

\item Empirical calibrations: consist of a calibration between a given emission line ratio
(e.g. $R_{23}$) and abundance estimates (e.g. O/H) derived from the $T_{\rm e}$-method (e.g. \citealt{2000A&A...362..325P, 2001A&A...369..594P, 2019ApJ...872..145J, 2021MNRAS.507..466D, 2025A&A...694A..61G}).
The main advantage of this method is to use reliable abundance values (by using the $T_{\rm e}$-method). However,  these calibrations are restricted to using abundance estimates in objects that have spectra with a high signal-to-noise ratio, low $Z$, and detectable auroral lines (e.g. \citealt{2006MNRAS.372..293H,2008MNRAS.383..209H,2007MNRAS.382..251D}).
   \item Theoretical calibrations: are based exclusively on photoionization models 
    (e.g. \citealt{1991ApJ...380..140M, 2002ApJS..142...35K, 2014MNRAS.443.1291D, 2024ApJ...977..187Z}), and the predicted emission line intensities (output) are calibrated with the metallicity considered in the models (input). The main advantage  of this method is to cover a wide range of nebular parameters (e.g. $Z$, ionization parameter, electron density among others). However, uncertainties in the physical processes considered in photoionization models due to assumptions such as electron temperature distribution, spectral energy distribution, geometry, atomic parameters, etc., introduce several uncertainties in this kind of calibration (e.g. \citealt{2013ApJS..208...10D, 2017MNRAS.466.4403N, 2017MNRAS.469.1036J, 2023ApJ...954..175Z, 2024arXiv240504598L, 2024ApJ...969L...5L, 2024A&A...684A..53B, 2025A&A...693A.215K}). 
    \item Semi-empirical calibrations: are based on a comparison
    between results of photoionization models and observational emission line ratios of a
    sample of objects.
    From this comparison, it is possible to infer a relation between the $Z$ values for the objects in the observational sample and their emission line intensity ratios (see  \citealt{2006A&A...447..863N, 2017MNRAS.467.1507C, 2020MNRAS.492.5675C, 2013MNRAS.432.2512D, 2019MNRAS.486.5853D}). The main advantage of this methodology is to minimize  the effects of the unrealistic
    nebular parameters in the models, constraining them through the observational data. Moreover, these methods do not require direct abundance estimates, which increases the potential sample size. Such calibrations are affected by photoionization model uncertainties and to the sample size. 
\end{itemize}

Another important factor is the spectral wavelength range necessary for the estimation of $Z$. For instance, in the optical range,
several auroral emission lines (e.g. [\ion{O}{iii}]$\lambda4363$, [\ion{N}{ii}]$\lambda5755$) can be measured, allowing for direct $Z$ estimates (i.e. to apply the $T_{\rm e}$-method) and making it possible to obtain empirical relations (e.g. \citealt{2016MNRAS.457.3678P, 2021MNRAS.507..466D}). Otherwise, in the case of the ultraviolet (UV)  range [$1000 \: < \: \lambda(\angstrom) \:  < \: 3000$]
direct abundance estimates are difficult or impossible to derive, requiring a combination with optical temperature estimates
(e.g. \citealt{2016ApJ...827..126B, 2024ApJ...971...87B, 2023ApJ...959..100I, 2023ApJ...955..112R, 2024MNRAS.535..881J, 2024arXiv240904625H, 2024ApJ...971...21H, 2024ApJ...969..148C, 2024MNRAS.529..781K, 2025ApJ...982...14H}).
An additional challenge in abundance estimates using UV emission lines is the limited number of hydrogen line measurements,
since the unique hydrogen line is Ly$\alpha$, a highly resonance line (e.g. \citealt{1995ApJ...441..507S, 2019MNRAS.486.2102H}). This result in an imprecise dust reddening correction (generally not considered or taken from the Balmer decrement, e.g. \citealt{2016ApJ...827..126B}) and in elemental abundance estimates relative to helium \citep[see][]{2022MNRAS.514.5506D}.
In summary, reliable $Z$ and metal abundance estimates from UV emission lines require both optical and UV spectroscopic data for the same source, which are available in the literature only for a few cases (e.g. \citealt{2016ApJ...827..126B, 2025MNRAS.540.1608D}).

The above caveats show that $Z$ estimates through only UV emission lines are subject to some uncertainties, as listed below. Photoionization models are generally used to estimate $Z$ from only UV emission lines, with the best emission line ratios used as $Z$ indicators being under discussion in the field of nebular astrophysics (e.g.
\citealt{2006A&A...447..863N, 2024ApJ...977..187Z}). \citet{2014MNRAS.443.1291D} suggested
the $C43$=(\ion{C}{iv}$\lambda1549$+\ion{C}{iiii}]$\lambda1909$)/\ion{He}{ii}$\lambda1640$ line ratio as
$Z$ indicator for Narrow Line Regions (NLRs) of AGNs. Theoretical calibrations relied on the $C43$ index, as well on other line
ratios (e.g. \ion{Si}{iii}]$\lambda1883, \lambda1892$/\ion{C}{iiii}]$\lambda1909$, \ion{N}{v}$\lambda1240$/\ion{He}{ii}$\lambda1640$; see \citealt{1992ApJ...391L..53H, 2006A&A...447..863N, 2024ApJ...977..187Z}),
require the knowledge of abundance relations. For instance, calibrations involving carbon emission lines are highly dependent on the C/O–O/H relation assumed in the photoionization models.
In particular, this relation was derived only for nearby \ion{H}{ii} regions
(e.g. \citealt{1999ApJ...513..168G, 2005ApJ...618L..95E, 2009ApJ...700..654E, 2020MNRAS.491.2137E, 2006ApJS..167..177D, 2020MNRAS.496.1051A, 2020ApJ...894..138S}),
not necessarily valid for high-$z$ objects (e.g. \citealt{2024MNRAS.535..881J}) and/or AGNs (e.g. \citealt{2023MNRAS.521.1556P}).

The scaling relations for the abundances of metals (e.g. C, N) as a function of total metallicity have been discussed in detail by \citet{2017MNRAS.466.4403N}.
These authors suggested using stellar abundance estimations to scale abundances with the total metallicity in nebular photoionization models. This approach has the advantage of producing abundance relations spanning over a wide range of
oxygen abundances, i.e.  $\rm 6.0 \: \lesssim \: 12+\log(O/H) 
\: \lesssim \: 9.0$ [$0.002 \: \lesssim \: (Z/\rm Z_{\odot}) \: \lesssim \: 2.0$] and, unlike most nebular abundances,
 stellar abundance relations are weakly affected by depletion of elements (e.g. C, O, Si)  onto dust grains
 (e.g. \citealt{1986PASP...98..995M, 1986Ap&SS.126..211O, 1989ApJ...338..162A, 1995ApJ...454..807S, 1995ApJ...449L..77G, 1995ApJ...439..793K, 2009ApJ...700.1299J, 2022MNRAS.512.2310G, 2023MNRAS.520.4345G, 2024A&A...690A.248M}).

Recently, \citep[][hereafter Paper~I]{2025MNRAS.540.1608D}, combined optical and UV data of AGNs (7 objects) together with 
a large sample of local \ion{H}{ii} regions (72 objects), and used the $T_{\rm e}$-method to derive a  C/O-O/H relation 
that showed deviations from those previously  proposed in the literature adopting  nebular (e.g. \citealt{2006ApJS..167..177D}) and stellar \citep{2017MNRAS.466.4403N} abundance estimates, 
in the sense that higher ($\sim 0.2$ dex) C/O abundances are derived for the very high
metallicity regime [$Z \: \gtrsim \: Z_{\odot}$ or $\rm 12+\log(O/H) \: \gtrsim \: 8.7$]. The use
of this new C/O-O/H relation to build photoionization models provides
$Z$ values from the $C43$ index for nearby AGNs in agreement with those via the $T_{\rm e}$-method, highlighting the reliability  of the derived abundance relation.
In this subsequent study, we use the C/O–O/H relation derived in Paper~I, combined with a large sample of type~2 AGN data, to provide a more reliable $C43$–$Z$ calibration for this class of active objects than previous ones \cite[see e.g.][]{2014MNRAS.443.1291D, 2019MNRAS.486.5853D}.
This new calibration is applied to a large sample of AGNs in a wide redshift range ($0 \: \lesssim \: z \:  \lesssim \: 12$), and the results are discussed in terms of the cosmic evolution of $Z$ and its relation with the stellar mass of the host galaxy as well as radio spectral indexes.
 The paper is organized as follows. In Section~\ref{meth} the methodology employed (observational data and photoionization models) to derive the new $C43$-$Z$ calibration
is presented. The results and  discussion are given in Sects.~\ref{res} and \ref{secdisc}, respectively.
The conclusion of the outcome is given in Sect.~\ref{conc}.
Throughout this paper, we
adopt the cosmological parameters by \citet{2021A&A...652C...4P}:
$\rm H_{0}$ = 67.4 km s$^{-1}$ Mpc$^{-1}$ and $\Omega_{\rm m}= 0.315$.

\section{Methodology}
\label{meth}
To obtain a semi-empirical $Z$ calibration, we use the C/O-O/H relation from Paper I as input for a grid of photoionization models constructed with the \textsc{Cloudy} code \citep{2017RMxAA..53..385F}. We then compare the model predictions with observational data on a $C43$ versus \ion{C}{iii}]$\lambda1909$/\ion{C}{iv}$\lambda1549$ diagram (see also \citealt{2025arXiv250109585A}). The procedure used to derive the calibration is described below.

\subsection{Photoionization models}
\label{mod}
We constructed a grid of photoionization models simulating the NLRs of AGNs using version 23.01 of the \textsc{Cloudy} code \citep{2017RMxAA..53..385F}. The input parameters for these models are described below.
\begin{itemize}
    \item Spectral Energy Distribution (SED): is parametrized
    by the $\alpha_{ox}$ slope \citep{1979ApJ...234L...9T}, defined by 
    \begin{equation}
    \label{eqaox}
\alpha_{ox}= \frac{\log [F(2\: {\rm kev})/F(2500 \: {\angstrom})]}{\log [\nu(2 \: {\rm keV})/\nu(2500 \:{\angstrom})]}, 
\end{equation}
where $F$ is the flux at the given frequence $\nu$. We assume $\alpha_{ox}=-1.1, -0.8$. 
The SED of an AGN depends mainly on the thermal emission from the surface of an accretion disk 
(e.g. \citealt{1982ApJ...254...22M,2019MNRAS.487.3884C}), on the thickness of the disk (e.g. \citealt{1989MNRAS.238..897L,2014SSRv..183...21B,2018ApJ...855..120T}) and on the  
electron scattering (e.g. \citealt{1987ApJ...321..305C,2021A&A...649A..87G}).
Photoionization models assuming these $\alpha_{ox}$ values are able to reproduce UV (e.g. \citealt{2018MNRAS.479.2294D}), 
optical (e.g. \citealt{2020MNRAS.492.5675C})  and
infrared (e.g. \citealt{2023A&A...679A..80C}) emission line intensity ratios. In \citet{2020MNRAS.492.5675C}, we found that models 
considering $\alpha_{ox} \: < \: -1.1$ result in emission line ratios
lower than the observational ones; thus, these values were not considered. 

A second ionization source by background cosmic rays
with an ionization rate (default value in the \textsc{Cloudy} code) of $\rm 2\: \times \: 10^{-16} \: s^{-1}$ \citep{2007ApJ...671.1736I} was adopted
in the models. This source influences the heating of the interstellar medium 
and ionization of ions with low ionization potential (e.g. $\rm N^{+}$, see
\citealt{2025A&A...693A.215K} and references therein). Also, a turbulence with velocity
equal to $\rm 100 \: km \: s^{-1}$  \citep{1992ApJ...389L..63F} was adopted in the models. Basically, turbulence affects the shielding and pumping of lines in the sense that fluorescent excitation  becomes more important for larger turbulent line widths.

    \item Metallicity ($Z$): the assumed values are
    $(Z/\rm Z_{\odot})= 0.1$, 0.2, 0.5, 1.0, 2.0, 3.0, and 4.0. They cover the $Z$ range derived for a large sample of AGNs with low (e.g. \citealt{2006MNRAS.371.1559G, 2019MNRAS.489.2652P, 2022MNRAS.513..807D, 2024MNRAS.529.4993L, 2024MNRAS.534.2723A}), intermediate (e.g. \citealt{2023ApJ...955..141C, 2024MNRAS.535..853J})  and high (e.g. \citealt{2023MNRAS.521.1556P, 2024MNRAS.535..881J})  redshifts, derived by using different methods
    (e.g. \citealt{2020MNRAS.492..468D, 2020MNRAS.496.3209D}). 
    As in the work by \citet{2006A&A...447..863N}, the models are dust free.
    \item Abundance relations: the abundances of all elements were linearly scaled with $Z$, with the exception of helium, nitrogen, and carbon, whose abundances were estimated from the relations derived through the $T_{\rm e}$-method  by \citet{2022MNRAS.514.5506D, 2024MNRAS.534.3040D, 2025MNRAS.540.1608D} and given by:  
    \begin{equation}
\label{eq1}
    \rm \log(N/O)=0.86 \times (x) - 8.39,  \: (x \: > \:8.0),
\end{equation}
\begin{equation}
    \rm \log(N/O)=-1.4, (x \: < \: 8.0),
\end{equation}
\begin{equation}
\label{eq2}
\rm 12+log(He/H) = 0.1215 \times (\rm x^{2})  -
1.8183 \times (\rm x) + 17.6732,
\end{equation}
and 
\begin{equation}
\label{coohrel}
\rm     \log(C/O)=0.41  \times  (x^{2}) 
    -6.07 \times (x) + 21.69,
\end{equation}
where $\rm x=12+\log(O/H)$.
\item Electron density ($N_{\rm e}$): three values of $N_{\rm e}$
were assumed: 100, 500 and 3000 cm$^{-3}$, which represents the 
range of electron density values derived from the [\ion{S}{ii}]$\lambda6716/\lambda6731$ line ratio for NLRs (e.g. \citealt{2017ApJ...850...40R, 2018A&A...618A...6K, 2018MNRAS.476.2760F, 2019A&A...622A.146M, 2020MNRAS.498.4150D, 2022MNRAS.516.1442I, 2024ApJ...960..108Z}).
    \item Ionization parameter ($U$): we consider the logarithm of the ionization
    parameter in the range of $-3.5 \: \leqq \: \log U \: \leqq\: -1.0$, with a step
    of 0.5 dex. Models assuming this range of values are able to reproduce
    observational data of AGNs with a wide range of ionization degree
    (e.g. \citealt{2020MNRAS.492.5675C, 2025arXiv250309267P}). The geometry of the models
    is plane parallel and the outermost radius is that where the electron temperature
    reaches 4000 K (default value of the \textsc{Cloudy} code). 
\end{itemize}
In total, 216 photoionization models were built, covering a wide  range of nebular parameters. 

\subsection{Observational data}
Our sample consists of the same observational data compiled from the literature by \citet{2014MNRAS.443.1291D, 2019MNRAS.486.5853D}, supplemented with 30 Type 2 Quasars ($2.0 \: \lesssim \: z \: \lesssim \: 4.0$) from \citet{2020MNRAS.495.4707S}, whose data were taken from the Sloan Digital Sky Survey III Baryon Oscillation Spectroscopic Survey (SDSS BOSS; \citealt{2013MNRAS.435.3306A}).

The data sample comprises 106 AGNs (redshift $0 \: 
\lesssim \: z \: \lesssim \: 4.0$) classified as Seyfert~2 (8 objects), Type~2 Quasars (36 objects), High-$z$ Radio Galaxies (61 objects), and  radio-quiet type-2 AGNs (1 object).
The stellar mass ($M_{\star}$) of the AGN host galaxies is in the range of $10 \: \lesssim \: \log (M_{\star}/\rm M_{\odot}) \: \lesssim \: 12$. However, for 51/106 galaxies of the sample, $M_{\star}$ is not available in the literature. 
The selection criteria adopted require that the objects have been classified as type~2 AGNs by the authors from whom the data were compiled, and have flux measurements of the \ion{C}{iii}]$\lambda1909$, \ion{C}{iv}$\lambda1549$, and \ion{He}{ii}$\lambda1640$ emission lines.  In general, the objects of our sample present Full Width at Half Maximum (FWHM)
lower than 1000 km s$^{-1}$, therefore, a low or negligible contribution of gas heating/ionization by shocks is expected (e.g. \citealt{1972ApJ...173L..13D, 1996ApJS..102..161D}).
The emission lines are not reddening corrected due to
two factors: ($i$) the line ratios considered have a near wavelength, being minimal
the effect of dust on the emission line
ratios (e.g. \citealt{1994ApJ...435..171K}) and ($ii$)
only one hydrogen line (Ly$\alpha$) is measured for  most objects. 

\begin{figure}
\includegraphics[angle=-90, width=0.4\textwidth]{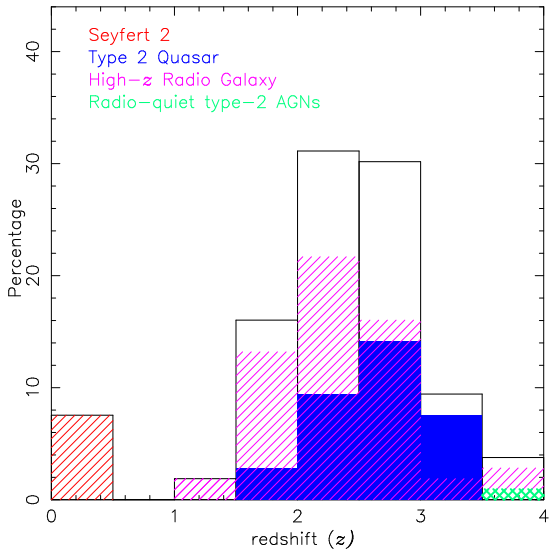}
 \caption{Redshift distribution of the  sample of objects (see Table~\ref{tab1}) used in the derivation of the new $C43$-$Z$ calibration. Distribution for the entire sample (106 objects) is represented by black solid line, while distinct AGN classes depicted by different colors, as indicated.}
\label{fig1}
\end{figure}

In Table~\ref{tab1}, identification,
redshift, values
of log($C43$) and log(\ion{C}{iii}]$\lambda1909$/\ion{C}{iv}$\lambda1549$),  low frequency radio luminosity measured at 325 MHz ($P_{325}$) 
and 1400 MHz ($P_{1400}$), are listed, when available. In Figure~\ref{fig1}, distributions of the AGN type in terms of the redshift are shown. We note that the
sample is dominated by high-$z$ radio galaxies ($\sim 57$ \%) and type~2 Quasars ($\sim 34$ \%)  in the range
of $1 \: \lesssim \: z \: \lesssim \: 4$.

In addition to the observational data of type~2 AGNs listed in Table~\ref{tab1}, we compiled information for 5 objects classified in the literature as possible AGNs at very high $z$, that is, in the range of $4.4 \: \lesssim \: z \: \lesssim \: 12.5$. The classification of these objects
was carried out by the authors from which the data were compiled through UV diagnostic diagrams 
(e.g. \citealt{2016MNRAS.456.3354F, 2018MNRAS.479.2294D})
or by the presence of high excitation lines, as, for instance, [\ion{Ne}{v}]$\lambda3426$ (see \citealt{2024arXiv241018193S}). 
These very high$-z$ objects were taken into account in the analysis of the results  (see below), but not in the calibration derivation due to the ambiguity of their ionization source (e.g. \citealt{2023AA...677A..88B}). The object identification, redshift,
log($C43$) and log(\ion{C}{iii}]$\lambda1909$/\ion{C}{iv}$\lambda1549$) of this auxiliary sample
are listed in Table~\ref{tab2}. Again, no reddening correction was performed for these data. 

\begin{figure*}
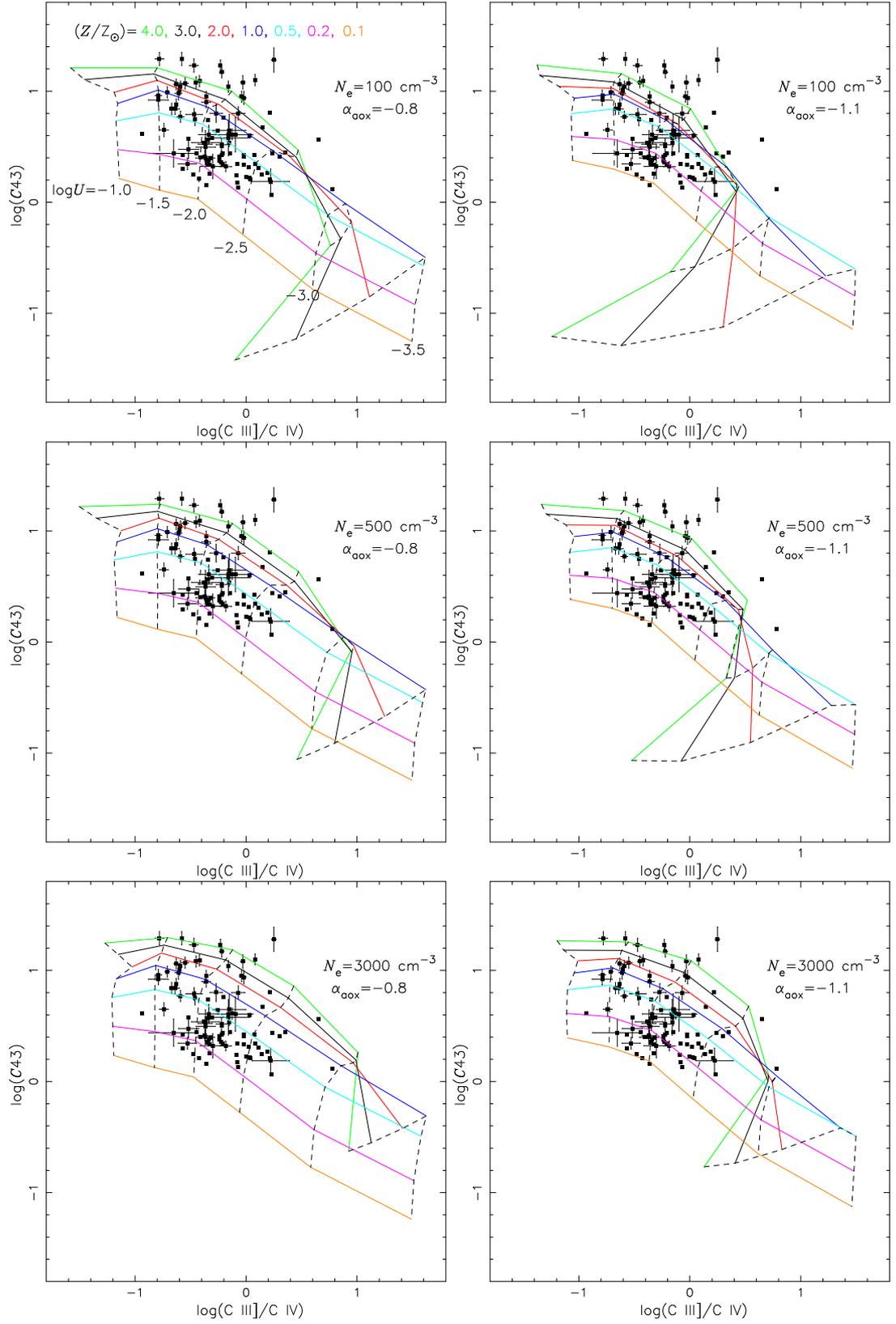

\includegraphics[angle=-90, width=0.4\textwidth]{fig_aox0.8_ne100.eps}
\includegraphics[angle=-90, width=0.4\textwidth]{fig_aox1.1_ne100.eps}
\includegraphics[angle=-90, width=0.4\textwidth]{fig_aox0.8_ne500.eps}
\includegraphics[angle=-90, width=0.4\textwidth]{fig_aox1.1_ne500.eps}
\includegraphics[angle=-90, width=0.4\textwidth]{fig_aox0.8_ne3000.eps} 
\includegraphics[angle=-90, width=0.4\textwidth]{fig_aox1.1_ne3000.eps}
\caption{$C43$=(\ion{C}{iv}$\lambda1549$+\ion{C}{iii}]$\lambda1909$)/\ion{He}{ii}$\lambda1640$
versus \ion{C}{iii}]$\lambda1909$/\ion{C}{iv}$\lambda$1549. Points represent observational narrow emission-line
intensity ratios listed in Table~\ref{tab1}. Solid lines connect photoionization model results
(see Sect.~\ref{mod}) with  same metallicity, as indicated in the left upper panel. 
Dashed lines connect models with same ionization parameter as indicated in the left upper panel.
In each panel, a grid of models assuming different $\alpha_{ox}$ (as defined in Eq.~\ref{eqaox})
and electron density ($N_{\rm e}$) values, are shown.}
\label{fig2}
\end{figure*}

\section{Results}
\label{res}

Fig.~\ref{fig2} presents diagrams of $\log(C43)$ versus $\log(\ion{C}{iii}]/\ion{C}{iv})$, incorporating both the observational data (listed in Table~\ref{tab1}) and the results of the photoionization models (see Sect.~\ref{mod}) that assume different nebular parameters, as indicated. The models successfully reproduce nearly all of the observational data, depending on the adopted nebular parameters. 
It should be noted that models with $(Z/\rm Z_{\odot}) \: > \: 1.0$ and  $\log U \: < \: -2.5$ result in the logarithm of the $C43$ and
\ion{C}{iii}]/\ion{C}{iv} line ratios overlapping other models with different parameters, making it impossible to derive accurate estimates. This effect arises from a combination of low electron temperatures ($T_{\rm e}\: \lesssim \: 7000 \: \rm K$) and a reduced number of ionizing photons, resulting in a weak emission line intensities.  Consequently, the validity range for nebular parameter estimations based on our calibrations is defined as $0.1 \: \lesssim \: (Z/\rm Z_{\odot}) \: \lesssim \: 4.0$ and $ -2.5 \: \lesssim \: \log U \: \lesssim \: -1.0$.  

Following \citet{2020MNRAS.492.5675C}, who derived a semi-empirical calibration for NLRs based on the relation between $N2$ = [\ion{N}{ii}]$\lambda6584$/H$\alpha$ and $Z$, we interpolated the model results in each diagram of Fig.~\ref{fig2} (for a similar methodology, see  \citealt{2006A&A...447..863N, 2009A&A...503..721M, 2018A&A...616L...4M, 2013MNRAS.432.2512D, 2021MNRAS.501.1370D, 2017MNRAS.467.1507C, 2020ApJ...898...26G, 2021MNRAS.505.2087K, 2023MNRAS.525.2087B}) deriving for each object, when possible,  sets of ($Z/\rm Z_{\odot}$, $\log U$) and their corresponding [$\log(C43)$, log(\ion{C}{iii}]/\ion{C}{iv})] values.
These types of diagrams (e.g., Fig.~\ref{fig2}) combine line ratios that mainly depend on $Z$ (e.g., $C43$, $R_{23}$) with those that primarily depend on the ionization degree (e.g., \ion{C}{iii}]/\ion{C}{iv}, [\ion{O}{iii}]/[\ion{O}{ii}]), thereby mitigating the effect of the ionization parameter on $Z$ estimates, and vice versa (e.g., \citealt{1991ApJ...380..140M, 2011MNRAS.415.3616D, 2016ApJ...816...23S, 2024PASA...41...99O}).

\begin{figure}
\includegraphics[angle=0, width=0.48\textwidth]{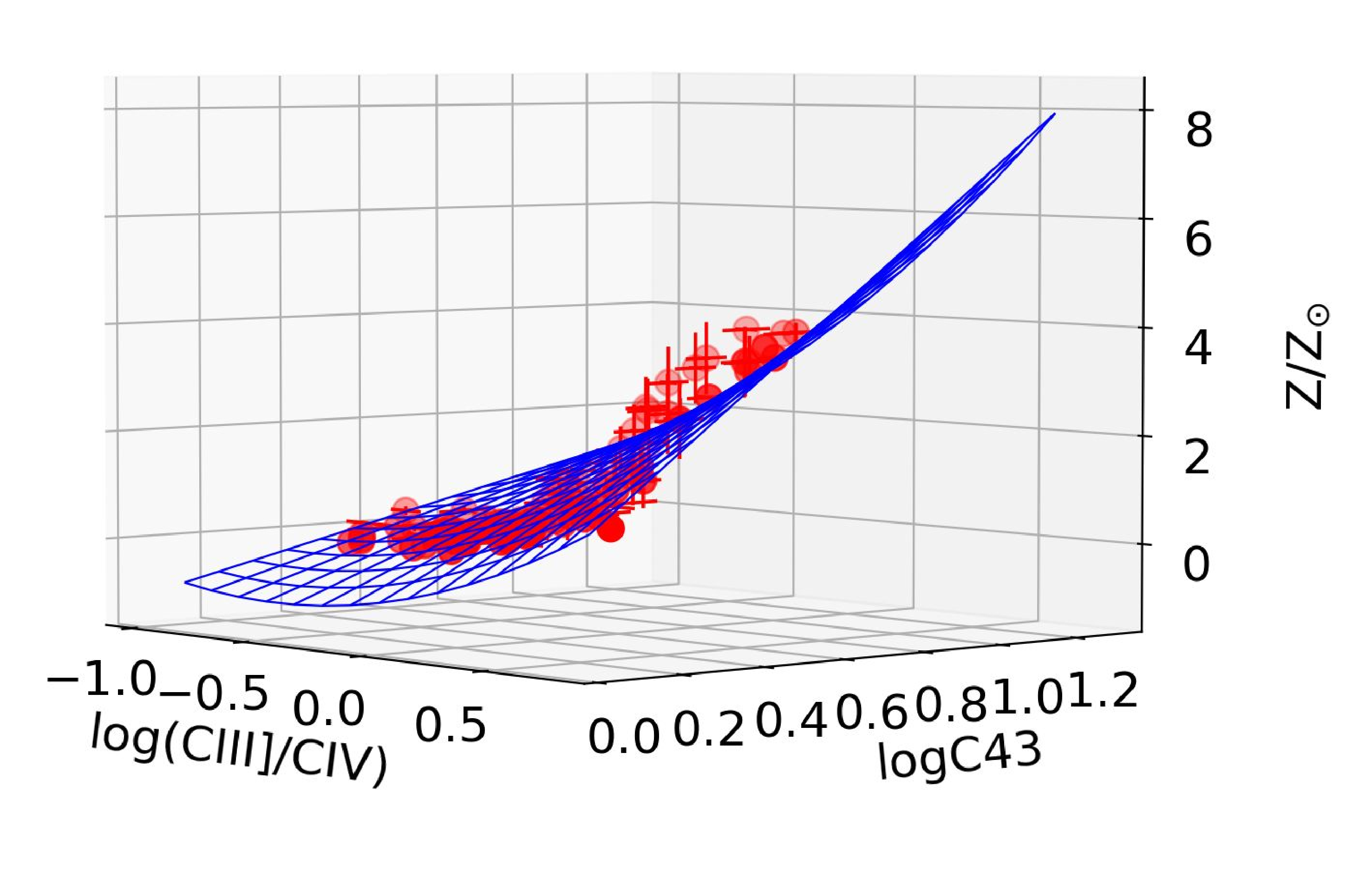}\\ 
\includegraphics[angle=-90, width=0.45\textwidth]{c34u.eps}
\caption{Upper panel: semi-empirical calibration between the metallicity
($Z/\rm Z_{\odot}$) and the logarithm of the $C43$=(\ion{C}{iv}$\lambda1549$+\ion{C}{iii}]$\lambda1909$)/\ion{He}{ii}$\lambda1640$
and \ion{C}{iii}]$\lambda1909$/\ion{C}{iv}$\lambda$1549 line ratios.
Points represent our $Z/\rm Z_{\odot}$ estimations (mean values) for the objects in our
sample (see Table~\ref{tab1})  obtained interpolating our photoionization model results presented in Fig.~\ref{fig2}. The surface represents the   fitting to the points given by Eq.~\ref{eqc43a}.
Lower panel: same as the
upper panel but for $\log U$ versus log(\ion{C}{iii}]$\lambda1909$/\ion{C}{iv}$\lambda1549$). The line represents the fitting to the points given
by Eq.~\ref{eqc3c4}.}
\label{figcv2}
\end{figure}

In Table~\ref{tab1}, the mean values for metallicity [$<(Z/\rm Z_{\odot})>$] and ionization parameter ($<\log U>$) calculated from the inferred values for each object in the panels of Fig.~\ref{fig2} are listed. The uncertainties of these parameters were calculated by propagating the individual errors derived for each object from the model interpolations (see also \citealt{2021MNRAS.501.1370D}). In general, the uncertainties are of the order of $\sim 10-20$\% in $Z$  estimates, reaching up to about 50\% for few cases, and $\sim0.1$ dex in $\log U$ (see also Table~1 in \citealt{2019MNRAS.486.5853D}).
In Fig.~\ref{figcv2}, upper panel,   
the resulting semi-empirical calibration  derived
by using the values listed in Table~\ref{tab1} is presented,
whose fitting to the points results in  
\begin{multline}
\label{eqc43a}
(Z/{\rm Z_{\odot}})=  (1.114 {\scriptstyle\pm 0.305}) \times w^{2} + (2.443 {\scriptstyle\pm 0.536}) \times wy \\ 
+(1.015 {\scriptstyle\pm 0.286}) \times w  + (4.167 {\scriptstyle\pm 0.206}) \times y -(0.928 {\scriptstyle\pm 0.110}), \\ 
\end{multline}
where $w$=log(\ion{C}{iii}]$\lambda1909$/\ion{C}{iv}$\lambda1549$) and
$y$=log($C43$). This calibration is valid for the line ratio intervals:
$0.0 \: \lesssim \: \log(C43) \: \lesssim \: 1.2$ and 
$-1.0 \: \lesssim \: \log(\ion{C}{iii}]/\ion{C}{iv}) \: \lesssim \: 1.0$.
In the same direction, we use the interpolated values from Fig.~\ref{fig2} and
derived a calibration for $\log U$ in terms of $w$=log(\ion{C}{iii}]/\ion{C}{iv}), given by 
\begin{equation}
\label{eqc3c4}
    \log U= (-1.021 \pm 0.025) \times w  -(2.282 \pm 0.010)
\end{equation}
and represented in the lower panel of Fig.~\ref{figcv2} by the red line.

\begin{table*}
\caption{Sample of objects compiled from the literature by \citet{2014MNRAS.443.1291D, 2019MNRAS.486.5853D} and 
used to obtain the new semi-empirical $C43$-$Z$ calibration.
Identification, redshift ($z$), log($C43$)=log[(\ion{C}{iv}$\lambda1549$+\ion{C}{iiii}]$\lambda1909$)/\ion{He}{ii}$\lambda1640$], log(\ion{C}{iii}]$\lambda1909$/\ion{C}{iv}$\lambda1549$), logarithm of the stellar mass of the host galaxy (in terms of $\rm M_{\odot}$),
 mean metallicity ($<Z/\rm Z_{\odot}>$) and ionization parameter 
($<\log U>$)  values derived from the photoionization model result interpolations
(Fig.~\ref{fig2}), luminosities  of 
Ly$\alpha$,  $P_{325}$ and $P_{1400}$ (see text) are listed when available.  Reference from which the radio luminosity values were obtained.}
\label{tab1}
\begin{tabular}{@{}lcccccccccc@{}}
\hline																										
Object                  &   $z$  &   $\log(C43)$   & log(\ion{C}{iii}]/\ion{C}{iv})  &  $\log\big(\frac{M_{\star}}{\rm M_{\odot}}\big)$  & $<Z/\rm Z_{\odot}>$  &  $<\log U>$  & log(Ly$\alpha$)	&  log($P_{325})$  & log($P_{1400})$ &  Refs.	     \\
\noalign{\smallskip}
\hline  																							 		   
                                                                             \multicolumn{11}{c}{Seyfert 2}                                                                  	      		 		   	                	      \\
\noalign{\smallskip}									     													 		   
NGC\,1068               &    0.004    &  $0.60\pm0.08$	&  $-0.33 \pm 0.09$   		  &  ---		      & $0.40\pm 0.03$       &   $-2.00 \pm 0.06$           &	 41.43  	&     ---	   &    30.26        &	1, 2  	      \\
NGC\,4507		&    0.012    &  $0.53\pm0.10$	&  $-0.36 \pm 0.12$ 		  &  ---		      & $0.32 \pm 0.04$	     &	 $-1.98 \pm 0.05$	    &	 41.41  	&     ---  	   &    29.36        &	1, 3  	      \\
NGC\,5506		&    0.006    &  $0.60\pm0.15$	&  $-0.09 \pm 0.15$  		  &  ---		      & $0.71 \pm 0.07$	     &	 $-2.21 \pm 0.07$	    &	   ---  	&     ---  	   &    29.46        &	1, 4  	      \\
NGC\,7674		&    0.029    &  $0.57\pm0.15$	&  $-0.15 \pm 0.19$ 		  &  ---		      & $0.50 \pm 0.07$      &	 $-2.17 \pm 0.08$	    &	 41.98  	&     ---  	   &    30.66        &	1, 2  	      \\
Mrk\,3                 	&    0.014    &  $0.52\pm0.05$	&  $-0.36 \pm 0.06$ 		  &  ---		      & $0.31 \pm 0.04$	     &	 $-1.98 \pm 0.05$	    &	 41.48  	&     31.14  	   &    30.71        &	1, 2  	      \\
Mrk\,573                &    0.017    &  $0.47\pm0.08$	&  $-0.51 \pm 0.09$  		  &  ---		      & $0.23 \pm 0.05$	     &	 $-1.81 \pm 0.08$	    &	 42.02  	&     ---  	   &    29.22        &	1, 2  	      \\
Mrk\,1388	        &    0.021    &  $0.49\pm0.08$	&  $-0.36 \pm 0.08$ 		  &  ---		      & $0.28 \pm 0.05$	     &	 $-1.99 \pm 0.05$	    &	   ---  	&     ---  	   &    29.02        &	1, 3  	      \\
MCG-3-34-64	        &    0.017    &  $0.32\pm0.10$	&  $-0.30 \pm 0.11$  		  &  ---		      & $0.20 \pm 0.04$	     &	 $-2.09 \pm 0.06$	    &	 41.59  	&     ---  	   &    30.28        &	1, 4  	      \\
\noalign{\smallskip}
\hline																								 		 
                                                                             \multicolumn{11}{c}{ Type 2 Quasar}                                                             	      		 		                        	     \\
\noalign{\smallskip}																						 		 
CDFS-031               &     1.603    &  $0.41\pm0.04$  &  $0.41 \pm 0.04$                & 11.43                     &   $1.45 \pm 0.46$    &  $-2.64 \pm 0.11$            &	   ---  	&     ---	   &  ---  	     &	1 	     \\
CDFS-057  	       &     2.562    &  $0.61\pm0.04$  &  $0.61 \pm 0.04$		  & 10.67    		      & 	---	     &	     ---	            &	  42.79         &     ---  	   &  ---  	     &	1 	     \\
CDFS-112a              &     2.940    &  $0.34\pm0.05$  &  $0.34 \pm 0.05$		  & ---      		      &   $0.82 \pm 0.07$    &	$-2.59 \pm 0.08$	    &	  42.65         &     ---  	   &  ---  	     &	1 	     \\
CDFS-153               &     1.536    &  $0.80\pm0.08$  &  $0.80 \pm 0.08$		  & ---      		      &    ---		     &	  ---	                    &	   ---	        &     ---  	   &  ---  	     &	1 	     \\
CDFS-531               &     1.544    &  $0.32\pm0.04$  &  $0.32 \pm 0.04$		  & 11.70    		      &   $0.74 \pm 0.07$    &	$-2.62 \pm 0.03$	    &	   ---	        &     ---  	   &  ---  	     &	1 	     \\
CXO\,52                &     3.288    &  $0.51\pm0.05$  &  $0.51 \pm 0.05$		  & ---      		      &    1.97		     &	$-2.63$	                    &	  43.28         &     ---  	   &  ---  	     &	1 	     \\
015307.0               &     2.33     &  $1.01\pm0.05$  &  $-0.60\pm 0.01$                & ---                       &   $1.74 \pm 0.48$    &  $-1.60\pm 0.06$             &     44.41         &     ---          &  ---            &  5	     \\
012403.3 	       &     2.60     &  $1.06\pm0.05$  &  $-0.63\pm 0.02$		  & --- 		      &   $2.16 \pm 0.56$    &	$-1.57\pm 0.05$		    &	  44.82 	&     ---	   &  ---	     &	5 	     \\
0130202.4	       &     2.64     &  $0.95\pm0.05$  &  $-0.04\pm 0.02$		  & --- 		      &   $3.24 \pm 0.63$    &	$-2.05\pm 0.08$		    &	  44.52 	&     ---	   &  ---	     &	5 	     \\
362507.4 	       &     2.87     &  $0.88\pm0.05$  &  $-0.63\pm 0.01$		  & --- 		      &   $0.78 \pm 0.07$    &	$-1.60\pm 0.07$		    &	  44.82 	&     ---	   &  ---	     &	5 	     \\
111507.9 	       &     2.81     &  $0.98\pm0.06$  &  $-0.61\pm 0.03$		  & --- 		      &   $1.40 \pm 0.40$    &	$-1.60\pm 0.07$  		    &	  44.73 	&     ---	   &  ---	     &	5 	     \\
344832.1 	       &     2.55     &  $1.08\pm0.05$  &  $-0.45\pm 0.02$		  & --- 		      &   $2.97 \pm 0.65$    &	$-1.72\pm 0.07$		    &	  44.55 	&     ---	   &  ---	     &	5 	     \\
450432.4               &     2.45     &  $0.99\pm0.05$	&  $-0.71\pm 0.02$		  & --- 		      &   $1.10 \pm 0.18$    &	$-1.51\pm 0.04$		    &	  44.84 	&     ---	   &  ---            &  5	     \\
382532.3 	       &     2.73     &  $1.17\pm0.05$	&  $-0.22\pm 0.02$		  & --- 		      &    3.66		     &	$-1.92$			    &	  44.64 	&     ---	   &  ---	     &	5 	     \\
031128.9 	       &     2.69     &  $1.07\pm0.05$	&  $-0.55\pm 0.03$		  & --- 		      &   $2.66\pm 0.66$     &	$-1.64\pm 0.06$		    &	  44.61 	&     ---	   &  ---	     &	5 	     \\
341538.0 	       &     2.40     &  $0.77\pm0.05$	&  $-0.59\pm 0.03$		  & --- 		      &   $0.49\pm 0.03$     &	$-1.68\pm 0.08$		    &	  44.48 	&     ---	   &  ---	     &	5 	     \\
401318.4 	       &     3.30     &  $0.96\pm0.05$	&  $-0.79\pm 0.03$		  & --- 		      &   $0.90\pm 0.05$     &	$-1.44\pm 0.07$		    &	  44.84 	&     ---	   &  ---	     &	5 	     \\
060027.4 	       &     2.27     &  $0.84\pm0.05$	&  $-0.64\pm 0.04$		  & --- 		      &   $0.62\pm 0.07$     &	$-1.62\pm 0.06$ 	    &	  44.50 	&     ---	   &  ---	     &	5 	     \\
072950.6               &     3.15     &  $0.90\pm0.05$	&  $-0.36\pm 0.04$		  & --- 		      &   $1.64\pm 0.69$     &	$-1.87\pm 0.10$             &	  44.89 	&     ---	   &  ---            &  5	     \\
074309.0 	       &     3.32     &  $0.84\pm0.05$	&  $-0.67\pm 0.02$		  & --- 		      &   $0.59\pm 0.08$     &	$-1.58\pm 0.06$		    &	  45.13 	&     ---	   &  ---	     &	5 	     \\
322649.6 	       &     2.71     &  $1.29\pm0.06$	&  $-0.58\pm 0.01$		  & --- 		      &    ---		     &	  ---			    &	  44.62 	&     ---	   &  ---	     &	5 	     \\
190534.8 	       &     2.54     &  $1.04\pm0.05$	&  $-0.59\pm 0.01$		  & --- 		      &   $2.12\pm 0.58$     &	$-1.60\pm 0.06$		    &	  44.98 	&     ---	   &  ---	     &	5 	     \\
260321.5 	       &     2.28     &  $0.65\pm0.06$	&  $-0.74\pm 0.04$		  & --- 		      &   $0.33\pm 0.05$     &	 $-1.50\pm 0.06$		    &	  44.46 	&     ---	   &  ---	     &	5 	     \\
233734.8 	       &     2.39     &  $1.04\pm0.05$	&  $-0.16\pm 0.01$		  & --- 		      &   $3.18\pm 0.46$     &	$-1.97\pm 0.07$		    &	  44.67 	&     ---	   &  ---	     &	5 	     \\
261026.2               &     3.07     &  $1.10\pm0.05$	&  $ 0.08\pm 0.01$		  & --- 		      &    ---		     &	  ---			    &	  44.83 	&     ---	   &  ---            &  5	     \\
005540.5 	       &     2.37     &  $1.09\pm0.05$	&  $-0.42\pm 0.01$		  & --- 		      &   $3.16\pm 0.66$     &	$-1.74\pm 0.07$		    &	  44.63 	&     ---	   &  ---	     &	5 	     \\
004550.5 	       &     2.72     &  $0.75\pm0.06$	&  $-0.46\pm 0.01$		  & --- 		      &   $0.52\pm 0.04$     &	$-1.85\pm 0.07$		    &	  44.21 	&     ---	   &  ---	     &	5 	     \\
023258.8 	       &     2.90     &  $0.80\pm0.06$	&  $-0.07\pm 0.06$		  & --- 		      &   $2.21\pm 0.69$     &	$-2.07\pm 0.08$		    &	  44.24 	&     ---	   &  ---	     &	5 	     \\
035715.5 	       &     2.22     &  $1.08\pm0.06$	&  $-0.03\pm 0.02$		  & --- 		      &   $3.76\pm 0.15$     &	$-2.04\pm 0.10$		    &	  44.34 	&     ---	   &  ---	     &	5 	     \\
000543.6 	       &     2.46     &  $1.28\pm0.11$	&  $ 0.25\pm 0.02$		  & --- 		      &    ---		     &	  ---			    &	  44.12 	&     ---	   &  ---	     &	5 	     \\
015218.3               &     3.18     &  $1.23\pm0.06$	&  $-0.47\pm 0.04$		  & --- 		      &    3.66		     &	$-1.70$			    &	  44.03 	&     ---	   &  ---            &  5	     \\
121119.9               &     3.14     &  $1.23\pm0.05$	&  $-0.23\pm 0.01$		  & --- 		      &    ---		     &	  ---			    &	  44.67 	&     ---	   &  ---	     &	5 	     \\
073116.0 	       &     3.05     &  $1.29\pm0.06$	&  $-0.78\pm 0.04$		  & --- 		      &    ---		     &	  ---			    &	  44.69 	&     ---	   &  ---	     &	5 	     \\
000848.4 	       &     2.22     &  $0.92\pm0.05$	&  $-0.79\pm 0.09$		  & --- 		      &    $0.78\pm 0.05$    &	$-1.44\pm 0.07$		    &	  44.14 	&     ---	   &  ---	     &	5 	     \\
010403.3 	       &     2.66     &  $0.94\pm0.06$	&  $-0.02\pm 0.02$		  & --- 		      &    $3.26\pm 0.62$    &	$-2.07\pm 0.08$ 	    &	  44.45 	&     ---	   &  ---	     &	5 	     \\
033833.5 	       &     2.76     &  $0.95\pm0.05$	&  $-0.36\pm 0.01$		  & --- 		      &    $2.17\pm 0.73$    &	$-1.83\pm 0.09$		    &	  44.42 	&     ---	   &  ---	     &	5 	     \\
\noalign{\smallskip}
\hline																								 		 
                                                                             \multicolumn{11}{c}{High-$z$ Radio Galaxy}                                                             	      		 		                	     \\
\noalign{\smallskip}																						 		 

USS\,1545-234           &    2.751    &  $0.34\pm0.01$  &   $-0.34 \pm 0.02$    	  & ---	     		      &     $0.19 \pm 0.04$  & $-2.04 \pm 0.06$		     &      ---         &    35.92   	   &  35.24          &	6, 7  \\  
USS\,2202+128           &    2.705    &  $0.53\pm0.01$  &   $-0.38 \pm  0.01$   	  & 11.62	     	      &     $0.31 \pm 0.04$  & $-1.96 \pm 0.06$ 	     &      ---         &    35.79   	   &  35.10          &	6, 7  \\  
USS\,0003-19            &    1.541    &      0.37	&      $-0.23$  	  	  & ---	     		      &     $0.25 \pm 0.07$  & $-2.15 \pm 0.06$		     &      ---         &    35.23   	   &  34.48          &	7 	     \\  
MG\,0018+0940           &    1.586    &      0.60	&  	0.03		  	  & ---	     		      &     $1.38 \pm 0.08$  & $-2.26 \pm 0.11$		     &      ---         &    35.36   	   &  34.84          &	7 	     \\  
MG\,0046+1102           &    1.813    &      0.42	&  	0.07		  	  & ---	     		      &     $0.53 \pm 0.05$  & $-2.42 \pm 0.05$		     &      ---         &    35.42   	   &  34.79          &	7 	     \\  
MG\,0122+1923           &    1.595    &      0.22	&  	0.00		  	  & ---	     		      &     $0.29 \pm 0.05$  & $-2.42 \pm 0.05$		     &      ---         &    35.24   	   &  34.77          &	7 	     \\  
USS\,0200+015           &    2.229    &      0.40	&      $-0.02$  	  	  & ---	     		      &     $0.42 \pm 0.03$  & $-2.34 \pm 0.06$		     &      ---         &    35.72   	   &  34.97          &	7 	     \\  
USS\,0211-122           &    2.336    &      0.40	&      $-0.40$  	  	  & $<11.16$	     	      &     $0.21 \pm 0.04$  & $-1.96 \pm 0.07$		     &     43.40        &    35.94   	   &  35.21          &	7 	     \\  
USS\,0214+183           &    2.130    &      0.42	&      $-0.22$  	  	  & --- 	     	      &     $0.30 \pm 0.05$  & $-2.15 \pm 0.05$		     &      ---         &    35.71   	   &  35.04          &	7 	     \\  
MG\,0311+1532           &    1.986    &      0.43	&      $-0.20$  	  	  & --- 	     	      &     $0.33 \pm 0.05$  & $-2.16 \pm 0.05$	             &      ---         &    35.37   	   &  34.81          &	7 	     \\  
\hline
\end{tabular}
\end{table*}

\begin{table*}
\setcounter{table}{0}
\caption{-$continued$}
\vspace{0.3cm}
\label{tab0}
\begin{tabular}{lcccccccccc}
\hline		 
\noalign{\smallskip}                                
Object                  &   redshift  &   $\log(C43)$          & log(\ion{C}{iii}]/\ion{C}{iv})        &  $\log(\frac{M_{\star}}{\rm M_{\odot}})$  & $<Z/\rm Z_{\odot}>$  &  $<\log U>$         &    log(Ly$\alpha$) & log($P_{325})$           & log($P_{1400})$  & 	Refs.		  \\
\noalign{\smallskip}																								     	 		                   	     
\hline 

                                                                                         \multicolumn{11}{c}{High-$z$ Radio Galaxy}                                                                                                                                  			    \\  		
\noalign{\smallskip}
TN\,J0121+1320          &    3.517    &  $0.21\pm0.02$         &   $0.03 \pm  0.01$         	       & 11.02      				   &	 $0.30 \pm 0.05$  & $-2.45 \pm 0.04$		  &	 ---	     &    35.83 	&  34.97	   & 6, 7 			    \\
TN\,J0205+2242          &    3.507    &  $0.39\pm0.04$         &   $-0.31\pm  0.05$         	       & 10.82      				   &	 $0.23 \pm 0.06$  & $-2.07 \pm 0.05$		  &	 ---	     &    35.88 	&  35.01	   & 6. 7 			    \\
MRC\,0316-257           &    3.130    &  $0.30\pm0.02$         &   $0.11 \pm  0.02$         	       & 11.20      				   &	 $0.42 \pm 0.02$  & $-2.49 \pm 0.04$		  &	 ---	     &     ---  	&   --- 	   & 6, 7 			    \\
USS\,0417-181           &    2.773    &  $0.26\pm0.03$         &   $0.19 \pm  0.04$         	       & ---	    				   &	 $0.44 \pm 0.01$  & $-2.57 \pm 0.02$		  &	 ---	     &    36.17 	&  35.39	   & 6, 7 		    \\
TN\,J0920-0712          &    2.758    &  $0.41\pm0.01$         &   $-0.23 \pm 0.01$         	       & ---	    				   &	 $0.29 \pm 0.06$  & $-2.14 \pm 0.05$		  &	 ---	     &    36.00 	&  35.05	   & 6, 7 			    \\
WN\,J1123+3141          &    3.221    &  $0.61\pm0.01$         &   $-0.93 \pm 0.06$       	       & $<11.72$   				   &	 $0.27 \pm 0.05$  & $-1.26 \pm 0.07$		  &	 ---	     &    35.99 	&  35.06	   & 6, 7 			    \\
4C\,24.28               &    2.913    &  $0.32\pm0.01$         &   $-0.18 \pm 0.02$       	       & $<11.11$   				   &	 $0.25 \pm 0.06$  & $-2.21 \pm 0.05$		  &	 ---	     &    ---		&  ---  	   & 6, 7 			    \\

BRL\,0310-150           &    1.769    &      0.57	       &      $-0.30$	            	       & ---	    				   &	 $0.38 \pm 0.05$  &  $-2.04 \pm 0.04$		  &	 ---	     &    36.12 	&  35.60	   & 7			    \\
USS\,0355-037           &    2.153    &      0.13	       &      $-0.06$	            	       & ---	    				   &	 $0.20 \pm 0.03$  &  $-2.38 \pm 0.05$		  &	43.60	     &    35.77 	&  34.99	   & 7			    \\
USS\,0448+091           &    2.037    &      0.44	       &       0.35	            	       & ---	    				   &	 $1.32 \pm 0.44$  &  $-2.58 \pm 0.10$		  &	43.58	     &    35.31 	&  34.64	   & 7			    \\
USS\,0529-549           &    2.575    &      0.56	       &       0.65	            	       & 11.46      				   &	     ---	  &   ---			  &	43.61	     &     ---  	&   --- 	   & 7			    \\
4C\,+41.17              &    3.792    &      0.60	       &      $-0.16$	            	       & 11.39      				   &	 $0.56 \pm 0.07$  &  $-2.16 \pm 0.06$		  &	44.31	     &    36.45 	&  35.66	   & 7			    \\
USS\,0748+134           &    2.419    &      0.34	       &      $-0.07$	            	       & ---	    				   &	 $0.34 \pm 0.04$  &  $-2.32 \pm 0.05$		  &	43.48	     &    35.72 	&  34.97	   & 7			    \\
USS\,0828+193           &    2.572    &      0.31	       &       0.02	            	       & $<11.60$   				   &	 $0.37 \pm 0.04$  &  $-2.41 \pm 0.05$		  &	43.87	     &    35.55 	&  34.84	   & 7			    \\
BRL\,0851-142           &    2.468    &      0.33	       &      $-0.32$	            	       &  ---	    				   &	 $0.20 \pm 0.04$  &  $-2.07 \pm 0.05$		  &	 ---	     &    36.00 	&  35.47	   & 7			    \\
TN\,J0941-1628          &    1.644    &      0.76	       &      $-0.20$	            	       &  ---	    				   &	 $1.10  \pm 0.48$ &  $-2.04 \pm 0.08$		  &	 ---	     &    35.71 	&  34.87	   & 7			    \\
USS\,0943-242           &    2.923    &      0.36	       &      $-0.22$	            	       &  11.22     				   &	 $0.25  \pm 0.06$ &  $-2.16 \pm 0.05$		  &	44.18	     &    36.10 	&  35.36	   & 7			    \\
MG\,1019+0534           &    2.765    &      0.25	       &      $-0.32$	            	       &  11.15     				   &	 $0.16  \pm 0.03$ &  $-2.07 \pm 0.05$		  &	42.74	     &    35.66 	&  35.19	   & 7			    \\
TN\,J1033-1339          &    2.427    &      0.57	       &      $-0.51$	            	       &  ---	    				   &	 $0.31  \pm 0.05$ &  $-1.80 \pm 0.07$		  &	43.67	     &    35.89 	&  35.02	   & 7			    \\
TN\,J1102-1651          &    2.111    &      0.20	       &	0.04	            	       &  ---	    				   &	 $0.29  \pm 0.04$ &  $-2.47 \pm 0.04$		  &	42.97	     &    35.57 	&  34.72	   & 7			    \\
USS\,1113-178           &    2.239    &      0.80	       &       0.21	            	       &  ---	    				   &	     2.58	  &    $-2.37$  		  &	43.40	     &    35.72 	&  35.08	   & 7			    \\
3C\,256.0               &    1.824    &      0.24	       &      $-0.08$	            	       &  ---	    				   &	 $0.25 \pm 0.05$  &  $-2.34 \pm 0.05$		  &	44.11	     &    36.15 	&  35.51	   & 7			    \\
USS\,1138-262           &    2.156    &      0.20	       &	0.21	            	       & $<12.11$   				   &  $0.40 \pm 0.02$	  &  $-2.59\pm 0.02$		  &	43.70	     &    36.34 	&  35.57	   & 7			    \\
BRL\,1140-114           &    1.935    &      0.50	       &      $-0.22$	            	       &  ---	    				   &  $0.37 \pm 0.05$	  &  $-2.12 \pm 0.06$		  &	 ---	     &    36.30 	&  35.60	   & 7			    \\
4C\,26.38               &    2.608    &      0.29	       &      $-0.56$	            	       &  ---	    				   &  $0.14 \pm 0.03$	  &  $-1.75 \pm 0.07$		  &	 ---	     &     ---  	&   --- 	   & 7			    \\
MG\,1251+1104           &    2.322    &      0.43	       &       0.23	            	       &  ---	    				   &  $0.84 \pm 0.08$	  &  $-2.51  \pm 0.05$  	  &	43.00	     &     ---  	&   --- 	   & 7			    \\
WN\,J1338+3532          &    2.769    &      0.06	       &       0.22	            	       &  ---	    				   &  $0.30 \pm 0.03$	  &  $-2.63  \pm 0.02$  	  &	44.07	     &    35.72 	&  34.97	   & 7			    \\
 MG\,1401+0921         	&     2.093   &  	0.17	       &     	   $-0.08$	               &    ---                     		   &   $0.21 \pm 0.04$    &  $-2.35  \pm 0.04$  	  &	 ---         &    35.53	        &  34.92           & 7			    \\
3C\,294               	&     1.786   &  	0.34	       &     	    0.07	  	       &    11.36    		    		   &   $0.43 \pm 0.03$    &  $-2.45  \pm 0.05$    	  &	 44.55       &   ---    	&   --- 	   & 7			    \\ 
USS\,1410-001         	&     2.363   &  	0.24	       &     	   $-0.47$	  	       &   $<11.41$  		    		   &   $0.13 \pm 0.03$    &  $-1.88  \pm 0.07$    	  &	 44.10       &   35.65  	&   35.00	   & 7			    \\ 
USS\,1425-148         	&     2.349   &  	0.15	       &     	   $-0.36$	  	       &   ---       		    		   &   $0.15 \pm 0.02$    &  $-2.08  \pm 0.02$    	  &	 43.96       &   35.65  	&   35.00	   & 7			    \\ 
USS\,1436+157         	&     2.538   &  	0.64	       &     	   $-0.25$	  	       &   ---       		    		   &   $0.50 \pm 0.07$    &  $-2.09  \pm 0.05$    	  &	 44.35       &   ---    	&   --- 	   & 7			    \\ 
3C\,324.0             	&     1.208   &  	0.42	       &     	   $-0.02$	  	       &   ---       		    		   &   $0.44 \pm 0.03$    &  $-2.34  \pm 0.06$    	  &	 ---         &   35.92  	&   35.34	   & 7			    \\ 
USS\,1558-003         	&     2.527   &  	0.36	       &     	   $-0.35$	  	       &    $<11.70$ 		    		   &   $0.20 \pm 0.04$    &  $-2.03  \pm 0.06$    	  &	 43.90       &   ---    	&   --- 	   & 7			    \\ 
BRL\,1602-174         	&     2.043   &  	0.42	       &     	   $-0.56$	  	       &   ---       		    		   &   $0.19 \pm 0.03$    &  $-1.75  \pm 0.08$    	  &	 ---         &   36.32  	&   35.67	   & 7			    \\ 
TXS\,J1650+0955       	&     2.510   &  	0.21	       &     	   $-0.42$	  	       &   ---       		    		   &   $0,13 \pm 0.03$    &  $-1.95  \pm 0.07$    	  &	 44.04       &   ---    	&   --- 	   & 7			    \\ 
8C\,1803+661          	&     1.610   &  	0.44	       &     	   $-0.44$	  	       &   ---       		    		   &   $0.22 \pm 0.04$    &  $-1.91  \pm 0.08$    	  &	  ---        &   34.55  	&   33.93	   & 7			    \\ 
4C\,40.36             	&     2.265   &  	0.33	       &     	   $-0.02$	  	       &    11.29    		    		   &   $0.36 \pm 0.04$    &  $-2.37  \pm 0.05$    	  &	  ---        &   36.24  	&   35.42	   & 7			    \\ 
BRL\,1859-235         	&     1.430   &  	0.24	       &     	     0.14	  	       &   ---       		    		   &   $0.39 \pm 0.02$    &  $-2.53  \pm 0.04$    	  &	  ---        &   36.29  	&   35.68	   & 7			    \\ 
4C\,48.48             	&     2.343   &  	0.38	       &     	   $-0.33$	  	       &   ---       		    		   &   $0.22 \pm 0.05$    &  $-2.05  \pm 0.06$    	  &	  ---        &   35.87  	&   35.20	   & 7			    \\ 
MRC\,2025-218       	&     2.630   &  	0.67	       &     	     0.14	  	       &    $<11.62$ 		    		   &   $1.78  \pm 0.36$   &  $-2.32  \pm 0.08$    	  &	 43.37       &   35.89  	&   35.25	   & 7			    \\ 
TXS\,J2036+0256       	&     2.130   &  	0.41	       &     	     0.30	  	       &   ---       		    		   &   $1.24  \pm 0.79$   &  $-2.53  \pm 0.10$    	  &	 43.38       &   35.86  	&   35.09	   & 7			    \\ 
MRC\,2104-242         	&     2.491   &  	0.53	       &     	    $-0.15$	  	       &    11.19    		    		   &   $0.44  \pm 0.04$   &  $-2.18  \pm 0.06$    	  &	 44.46       &   36.26  	&   35.41	   & 7			    \\ 
4C\,23.56             	&     2.483   &  	0.34	       &     	    $-0.21$	  	       &    11.59    		    		   &   $0.25  \pm 0.06$   &  $-2.18  \pm 0.06$    	  &	 43.61       &   ---    	&   --- 	   & 7			    \\ 
MG\,2121+1839         	&     1.860   &  	0.74	       &     	    $-0.34$	  	       &    ---      		    		   &   $0.64  \pm 0.06$   &  $-1.97  \pm 0.06$    	  &	 ---         &   35.32  	&   34.73	   & 7			    \\ 
USS\,2251-089         	&     1.986   &  	0.56	       &     	    $-0.34$	  	       &    ---      		    		   &   $0.36  \pm 0.04$   &  $-2.00  \pm 0.04$    	  &	 ---         &   35.66  	&   35.07	   & 7			    \\ 
MG\,2308+0336       	&     2.457   &  	0.44	       &     	    $-0.14$	  	       &    ---      		    		   &   $0.37  \pm 0.05$   &  $-2.22  \pm 0.06$    	  &	 43.16       &   35.89  	&   35.32	   & 7			    \\ 
4C\,28.58           	&     2.891   &  	0.11	       &     	     0.77	  	       &    11.36    		    		   &   $0.90  \pm 0.06$   &  $-2.94  \pm 0.03$    	  &	 ---         &   36.30  	&   35.44	   & 7			    \\

MP \,J0340-6507       	&     2.289   &     $0.18 \pm0.11$     &     	$0.22  \pm 0.16$   	       &    --- 		   		   &   $0.39  \pm 0.02$   &  $-2.60  \pm 0.02$    	  &	 42.22       &   ---    	&    ---	   & 8			    \\ 
TN\,J1941-1951        	&     2.667   &     $0.43 \pm0.12$     &     	$-0.65 \pm 0.22$ 	       &    --- 		   		   &   $0.18  \pm 0.03$   &  $-1.62  \pm 0.08$    	  &	 43.35       &   ---    	&    ---	   & 8			    \\ 
MP\,J2352-6154        	&     1.573   &     $0.39 \pm0.06$     &     	$-0.36 \pm 0.08$  	       &    --- 		   		   &   $0.21  \pm 0.05$   &  $-2.01  \pm 0.06$  	  &	 ---         &   ---    	&    ---	   & 8			    \\ 
\hline																										     	 		      
                                                             \multicolumn{11}{c}{Radio-quiet type-2 AGNs}                           		   						 		                                                          		   \\  
							     																			     	 		   
COSMOS\,05162           &    3.524    &     $0.78 \pm0.05$     &        $-0.46  \pm 0.07$              &    10.50                  		   &  $0.62 \pm 0.08$	  &  $-1.84 \pm 0.08$	           &	 43.96       &   ---             &    ---          & 9             	   \\
\hline																													 
\end{tabular}
\begin{minipage}[c]{2\columnwidth}
References--- (1) \citet{2006A&A...447..863N} , (2) \citet{2002AJ....124..675C}, (3) \citet{1998AJ....115.1693C}, (4) \citet{2014MNRAS.440..696A}, (5) \citet{2020MNRAS.495.4707S}, 
(6) \citet{2009A&A...503..721M}, (7) \citet{2000A&A...362..519D}, (8) \citet{2007MNRAS.378..551B}, (9) \citet{2018A&A...616L...4M}. 
\end{minipage}
\end{table*}

\begin{table*}
\caption{Sample of very high-$z$ objects classified as AGNs. Emission line ratios
are not reddening corrected and were compiled from the literature, as indicated.
These data were not included in the derivation of the  calibration (see Fig.~\ref{figcv2}) but were included as auxiliary data along the analysis. Metallicity and ionization parameter were derived by using calibrations represented
by Eqs.~\ref{eqc43a} and \ref{eqc3c4}, respectively.}
\label{tab2}
\begin{tabular}{lcccccc}
\hline
Object                  &   redshift  &   $\log(C43)$   & log(\ion{C}{iii}]/\ion{C}{iv})  & $Z/\rm Z_{\odot}$  &  $\log U$ &    Reference                         \\
\noalign{\smallskip}
\hline  
 106462                 &  8.51       &    0.33         &      $-0.14$                    &     0.21           & $-2.13$   &  \citet{2024arXiv240912232T}          \\
A744-45924              &  4.47       &    1.11         &       0.38                      &     $>4$           & $-2.66$   &  \citet{2024arXiv241204557L}           \\
GHZ2                    &  12.33      &    1.11         &      $-0.45$                    &     2.24          & $-1.82$   &  \citet{2024ApJ...972..143C}           \\
GN-z11                  &  10.60      &    0.56         &        0.33                     &     2.31          & $-2.61$   &  \citet{2023AA...677A..88B}            \\
GS-3073                 &  5.55      &    0.39          &        0.31                     &     1.41          & $-2.59$   &  \citet{2024MNRAS.535..881J}           \\
 \hline
\end{tabular}
\end{table*}

\section{Discussion}
\label{secdisc}

\subsection{Comparison with other calibrations}

 \begin{table*}
\caption{Summary of the calibrations used to  compare (Fig.~\ref{figcomp}) with our new
$C43$-$Z$ calibration (Eq.~\ref{eqc43a}). Columns: (1) Reference, (2) metallicity index used as $Z$ tracer, (3) calibration identification (ID) used along the text, (4) $Z$ validity range, (5) type of calibration.}
\label{tabcal}
\centering
\begin{tabular}{@{}lcccc@{}} 
\hline
Reference                    &  Index Name  &  ID            &         Validity                                               & Type            \\
\hline
                            &  \multicolumn{4}{c}{Ultraviolet}                                            \\
                   
\citet{2019MNRAS.486.5853D} &   ${C43\,}^a$                                &  D19           &  $0.3 \: \lesssim \: (Z/\rm Z_{\odot}) \: \lesssim \: 2.0$     & Semi-empirical  \\          
\citet{2024ApJ...977..187Z} &   ${C43\,}^a$                                &  Z24           &  $0.03 \: \lesssim \: (Z/\rm Z_{\odot}) \: \lesssim \: 6.5$       & Theoretical     \\
                            &  \multicolumn{4}{c}{Optical}                                                                                                               \\
\citet{2020MNRAS.492.5675C} &   ${N2\,}^b$                                 &  C20           & $0.3 \: \lesssim \: (Z/\rm Z_{\odot}) \: \lesssim \: 2.0$      & Semi-empirical   \\
\citet{2021MNRAS.507..466D} &   ${R_{23}\,}^c$                             &  D21           & $0.3 \: \lesssim \: (Z/\rm Z_{\odot}) \: \lesssim \: 2.0$      & Empirical        \\
\citet{1998AJ....115..909S} &    ${N2\,}^b$                                &  SB98          & $0.5 \: \lesssim \: (Z/\rm Z_{\odot}) \: \lesssim \: 5.0$      & Theoretical       \\
\hline
\end{tabular}\\

\begin{minipage}[c]{0.65\textwidth}
    $^{a}C43=(\ion{C}{iv}\lambda1549+\ion{C}{iii}]\lambda1909)/\ion{He}{ii}\lambda1640$; 
   $^{b}N2=[\ion{N}{ii}]\lambda6584/\rm H\alpha$;\\
    $^{c}R_{23}=([\ion{O}{ii}]\lambda3727+[\ion{O}{iii}]\lambda4959+\lambda5007)/\rm H\beta$
\end{minipage}
\end{table*}

Recently, metallicity and element abundance (O/H, Ne/H, S/H, Ar/H, C/H) estimates have been conducted using the $T_{\rm e}$-method for a relatively large sample of AGNs, comprising approximately 150 objects (see, e.g., \citealt{2020MNRAS.496.2191F, 2020MNRAS.496.3209D, 2022MNRAS.514.5506D, 2023MNRAS.521.1969D, 2024MNRAS.534.3040D, 2025MNRAS.540.1608D, 2021MNRAS.508..371A, 2021MNRAS.508.3023M, 2024MNRAS.535..881J}). In particular, \citet{2025MNRAS.540.1608D}  derived, for the first time using the $T_{\rm e}$-method, a C/O–O/H relation (i.e. Eq.~\ref{coohrel}) representative of NLRs. This new relation was obtained by combining direct estimates from both SFs and AGNs, following a method similar to that developed for other elements (e.g. \citealt{2020MNRAS.496.2191F, 2021MNRAS.508.3023M, 2021MNRAS.508..371A, 2023MNRAS.521.1969D, 2024MNRAS.534.3040D}). The use of this abundance relation in photoionization models yields, in principle, more reliable $Z$ calibrations for AGNs based on carbon lines, compared to those relying on other abundance relations that assume SF (e.g. \citealt{2022MNRAS.513.5134N}) and stellar (e.g. \citealt{2024ApJ...977..187Z}) abundances or fixed C/O values (e.g. \citealt{2006A&A...447..863N, 2016MNRAS.456.3354F, 2014MNRAS.443.1291D, 2019MNRAS.486.5853D}).

As usual, any new calibration must be compared with previous calibrations (e.g., \citealt{2024ApJ...977..187Z}). Thus, we compare the metallicity estimates from our new calibration 
(Eq.~\ref{eqc43a}) with those derived from other UV and optical calibrations. A detailed description of the calibrations used in this comparison is presented by \citet{2020MNRAS.492..468D, 2025MNRAS.540.1608D} and a summary is provided in Table~\ref{tabcal}. In addition, we compare the $Z$ values obtained via our $C43$–$Z$ calibration with those derived from the $T_{\rm e}$-method. Due to the requirement [e.g. $\rm (S/N) \:  \gtrsim \: 2$] for reliable measurements of auroral lines (e.g., [\ion{O}{iii}]$\lambda4363$), this comparison was performed only for the six nearby Seyfert~2 galaxies in our sample.  Regarding the application of the optical calibrations, the comparison was possible for only a few (6 through the $T_{\rm e}$-method and 7 through the $N_2$ or $R_{23}$ indexes)  objects in our sample. The optical emission line intensities (not shown) used in the estimates obtained using the $T_{\rm e}$-method and from optical calibrations were taken from \citet{2025MNRAS.540.1608D}. For the UV calibrations, we used the emission-line intensity ratios  listed in Table~\ref{tab1}.

\begin{figure*}
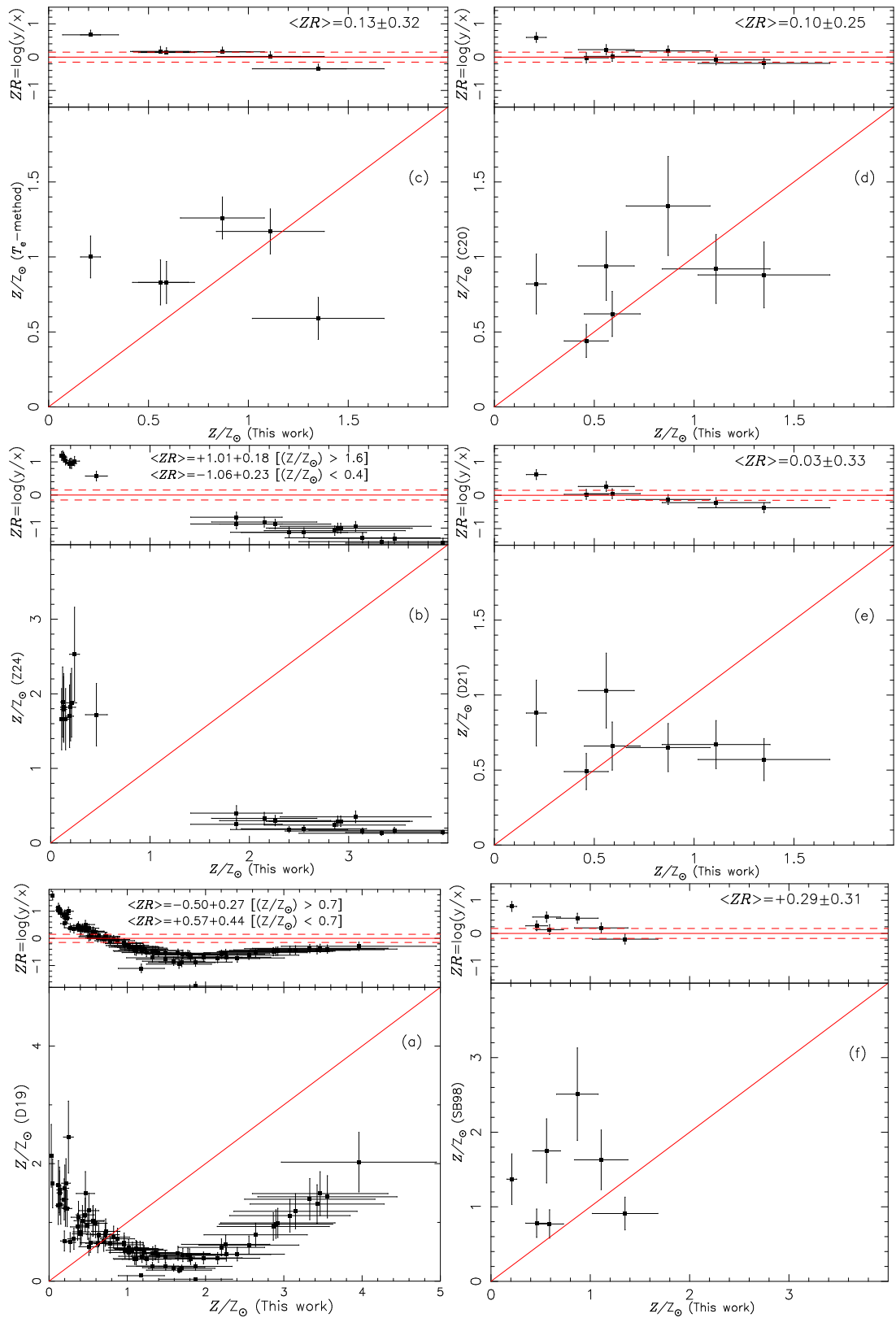

\includegraphics[angle=-90, width=0.4\textwidth]{compz_newd.eps} 
\includegraphics[angle=-90, width=0.4\textwidth]{compz_newa.eps} 
\includegraphics[angle=-90, width=0.4\textwidth]{compz_newg.eps} 
\includegraphics[angle=-90, width=0.4\textwidth]{compz_newb.eps} 
\includegraphics[angle=-90, width=0.39\textwidth]{compz_new.eps} 
\includegraphics[angle=-90, width=0.4\textwidth]{compz_newh.eps} 
\caption{Bottom part of panels: comparison between $Z$ values (in relation to the solar one) derived from our new $C43$-$Z$ calibration, represented by
Eq.~\ref{eqc43a} and D25, and the ones listed in Table~\ref{tabcal}. Points represent metallicity  estimates for the object in our sample (see Table~\ref{tab1}) through distinct calibrations, as indicated. The red lines represent the equality between the estimates. Upper part of panels: logarithm of the ratio between the metallicity estimates $(ZR)$ compared in the bottom part of the corresponding panel versus the estimations from our new $C43$-$Z$ calibration (Eq.~\ref{eqc43a}). Solid lines represent the equality between the estimates, whilst dashed lines the ratio uncertainty of them. The mean value of $ZR$ is indicated by $<ZR>$.}
\label{figcomp}
\end{figure*}

In Fig.~\ref{figcomp}, panels (a) and (b), the comparison between the $Z$ estimates from the new calibration (Eq.~\ref{eqc43a}) and those from  \citet{2019MNRAS.486.5853D} and \citet{2024ApJ...977..187Z} UV calibrations, respectively, is shown. We can see that the new $C43$ calibration produces, for the high-metallicity regime [$(Z/\rm Z_{\odot}) \: \gtrsim \: 0.7$],  higher $Z$ values than those derived via the other UV calibrations, while the opposite result is found for the low-metallicity regime. In the upper part of the panels (a) and (b) of Fig.~\ref{figcomp}, the logarithm of the ratio $ZR$ between the estimates as a function of $Z/\rm Z_{\odot}$ estimated using the present calibration is shown. The mean discrepancies ($<ZR>$) are about $-0.5$ and $0.6$ dex for the low and high-metallicity regime, respectively, when compare our estimations with those using the calibration by \cite{2019MNRAS.486.5853D}, while they are about $-1.0$ and 1.0 for the different metallicity regimes, respectively, when we performed the same comparison using the calibration by \cite{2024ApJ...977..187Z}.

\citet{2025MNRAS.540.1608D} proposed that the above discrepancy is mainly due to the inappropriate C/O–O/H relation assumed in the photoionization models used to derive the calibrations. If this assertion is right, the $Z$ values from this new calibration
(Eq.~\ref{eqc43a}) must be closer to those by the direct estimates. In Fig.~\ref{figcomp}, panel (c), our $Z$ estimates are compared with those obtained using the $T_{\rm e}$-method. Although this comparison is only possible for a few (6) objects, a good agreement is found with $<ZR> \sim 0.1$ dex, a value similar to the typical uncertainty in direct estimates due to the errors in emission-line measurements (e.g. \citealt{2003ApJ...591..801K, 2008MNRAS.383..209H}).
As an additional test, in panels (d), (e) and (f) of Fig.~\ref{figcomp}, we compare $Z$ derived from our calibration (Eq.~\ref{eqc43a}) with those obtained via optical calibrations by \citet{2020MNRAS.492.5675C}, \citet{2021MNRAS.507..466D} and \citet{1998AJ....115..909S}, respectively. We note very good agreement between our new estimates and those derived using  \citet{2020MNRAS.492.5675C} and \citet{2021MNRAS.507..466D} calibrations, with $<\!ZR\!>$ lower than $\sim 0.1$ dex. The \citet{1998AJ....115..909S} calibration results in somewhat higher ($<\!ZR\!>\sim 0.3$ dex) $Z$ values. This calibration adopts $N2=[\ion{N}{ii}]\lambda6584/\rm H\alpha$ as the $Z$ tracer and is based on photoionization models that assume an old N/O–O/H relation derived for nuclear starbursts \citep{1994ApJ...429..572S}, which can not necessarily be representative of AGNs (e.g. \citealt{2020MNRAS.496.2191F, 2024MNRAS.534.3040D, 2024MNRAS.535..881J}). In any case, $Z$ values derived from our new $C43$ calibration are consistent with those derived from the $T_{\rm e}$-method and most optical calibrations. It is worth mentioning that a large number of both UV and optical $Z$ estimates will be needed to confirm this result.

\subsection{Metallicity implications}

\subsubsection{Cosmic metallicity evolution}

The metallicity evolution of galaxies is connected to several physical processes operating over their lifetimes, such as mergers, interactions, supernova feedback, gas inflow/outflow, and the environment in which these objects are located  (e.g., \citealt{1990AJ.....99.1740H, 2007MNRAS.382..801M, 2011MNRAS.416...38K, 2011MNRAS.416.1354D, 2012ApJ...748...48R, 2014A&A...563A..58T, 2017MNRAS.468.1881W, 2009MNRAS.396.1257E, 2023ApJ...954...38K, 2023ApJ...956L..40W, 2025ApJ...980...12S}). These complex processes have a strong influence on the metallicity evolution of galaxies, producing a large scatter in $Z$ at a given redshift and/or for galaxies with similar mass
(e.g., \citealt{2018A&A...616L...4M, 2024NatAs...8..368P, 2025A&A...694A..18P, 2024A&A...690A.397V}). 

Recent galaxy $Z$ estimations performed by \citet{2025MNRAS.539.2463S} using the JWST
showed the existence of star-forming galaxies with  metallicity or elemental abundance (e.g. N, C) values higher than those predicted  by standard chemical evolution models (see \citealt{2024MNRAS.527.8193D} and references therein). Recent analysis by \citet{2025A&A...697A..96R} showed that
the discrepancy between observational estimated and model predicted values of $Z$ and elemental abundances can be attributed to the fact that standard chemical evolution models of galaxies assume inappropriate star formation rates and/or normal (e.g. \citealt{1955ApJ...121..161S, 2001MNRAS.322..231K})
stellar Initial Mass Function (IMF), being not
necessary to assume peculiar nucleosynthesis from population III stars (see also \citealt{2025arXiv250512505J, 2025arXiv250412584G, 2025arXiv250311457N, 2025arXiv250212091I}).

\begin{figure}
\includegraphics[angle=-90, width=0.4\textwidth]{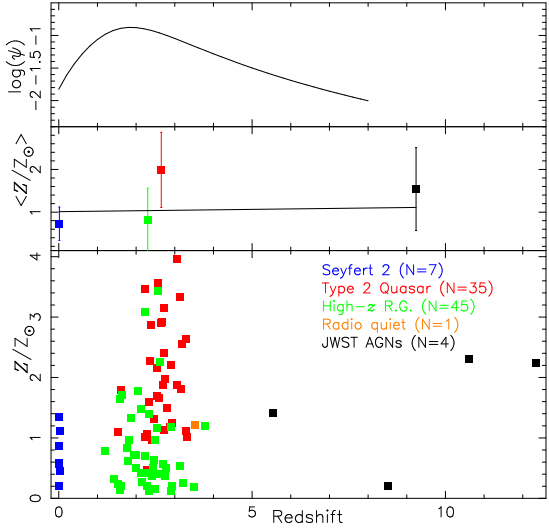}
 \caption{Bottom panel: metallicity (in relation to the solar metallicity value) versus the redshift for the objects in our sample (Table~\ref{tab1}). $Z$ values are derived assuming our new $C43$-$Z$ calibration (Eq.~\ref{eqc43a}).
AGN classes are represented by different colors, as indicated. Black point represent 
JWST estimates for the high-z AGNs listed in Table~\ref{tab2}. N represents the number of objects. Middle panel:
mean metallicity for the distinct AGN classes versus their mean redshift. The line represents a linear fitting (coefficients not shown) to the points.
Upper panel: logarithm of the cosmic star formation ($\psi$)
history derived by \citet{2014ARA&A..52..415M}.}
\label{figcos}
\end{figure}

Due to the high luminosity of AGNs, most of their narrow and broad emission lines are observed with  $\rm (S/N) \: > \:3$, thus permitting the estimation of $Z$ over a wide redshift range
(e.g. \citealt{2006A&A...447..157N, 2006A&A...447..863N, 2009A&A...494L..25J, 2010MNRAS.407.1826S, 2009A&A...503..721M, 2012A&A...542L..34N, 2016MNRAS.456.3354F, 2018ApJ...856...89T, 2019MNRAS.489.2652P, 2019A&A...626A...9M, 2020ApJ...898...26G, 2022A&A...666A.115P, 2022MNRAS.513..807D, 2022ApJ...929...51T, 2023ApJ...955..141C, 2024ApJ...977..187Z}). Our new calibration provides an opportunity to revise the
$Z$ evolution in AGNs, particularly in their NLRs.
In this context, in Fig.~\ref{figcos}, bottom panel, $Z$ estimates for our AGN sample (see Table\ref{tab1}), combined
with those for high-$z$ objects (see Table~\ref{tab2}), versus redshift are shown. 
In Fig.~\ref{figcos}, middle panel, we present the mean metallicity and redshift values for our 
sample, where the points represent $Z$ estimates for each class of AGN, 
as indicated. Finally, in the upper panel of Fig.~\ref{figcos}, we present the cosmic star formation ($\psi$)
history derived by \citet{2014ARA&A..52..415M} through ultraviolet and
infrared observational data. We can see that:
\begin{itemize}
    \item The highest $Z$ values (around $\sim 4 \: \rm Z_{\odot}$) are derived for Type~2 quasars and high-$z$ radio galaxies located in the redshift interval $2 \: \lesssim \: z \: \lesssim \: 3$. Interestingly, this redshift range is similar to that derived for the maximum value of $\psi$, according to the results by \citet{2014ARA&A..52..415M} (see also \citealt{2017A&A...602A...5N, 2024A&A...683A.174W, 2025A&A...697A..46G, 2025ApJ...984..117B}, among others). This agreement can be due to the possible link among black hole accretion rate (BHAR), star formation rate (SFR) and galaxy stellar mass ($M_{\star}$), resulting in a strong  enrichment of metals in the ISM (e.g. \citealt{2017ApJ...842...72Y, 2024ApJ...964..183Z, 2024A&A...683A.160M}). 
    However, we strongly emphasize that the agreement between the maximum $Z$ and $\psi$ values observed in Fig.~\ref{figcos} must be confirmed
    by additional  estimates of $Z$ in AGNs located in a more uniform range of redshifts.  Also, Type~2 quasars and High-$z$ radio  
    galaxies usually are hosted in more massive galaxies compared to those of
    Seyfert~2 nuclei, which could yield an observational bias
    in the above result, especially if $Z_{\rm AGN}$ is related to the mass of the host galaxy (see below).
    \item  The few possible $Z$ estimates for AGNs observed by the JWST are similar to those for objects in the local universe and at $1.0 \: \lesssim \: z \: \lesssim 4.0$. This result is consistent with that obtained by \citet{2006A&A...447..863N}, which indicates that AGNs had already undergone a major episode of metal enrichment in the early ($z \: > \: 5$) epoch of the Universe (see also \citealt{2003ApJ...596L.155M, 2007ApJ...669...32K, 2009A&A...503..721M, 2011ApJ...739...56D, 2014ApJ...790..145D, 2020ApJ...905...51S, 2020ApJ...898..105O, 2022MNRAS.513.1801L, 2024ApJ...975..214J}).
\end{itemize}

\subsubsection{Ionization parameter}

The ionization parameter $U$ characterizes the interaction between the ionizing source and the ionized gas (e.g., \citealt{2016ApJ...816...23S}) and it
represents the degree of excitation of the gas in a given object, reflecting the recombination rate at the face of a cloud exposed to radiation \citep{2006LNP...693...77P}.  
As it is already known, it is defined by
\begin{equation}
\label{equ1}
U = \frac{Q(\rm H)}{4\pi \: n({\rm H}) \: r^{2} \: \rm c},
\end{equation}
where $Q(\rm H)$ is the number of hydrogen-ionizing photons emitted by the central object,
$r$ is the separation [cm] between the center of the ionizing source and the illuminated face of the cloud,
$n({\rm H})$ is the total hydrogen density (ionized, neutral, and molecular), and "c" is the speed of light [cm $\rm s^{-1}$]. 

\begin{figure}
\includegraphics[angle=-90, width=0.4\textwidth]{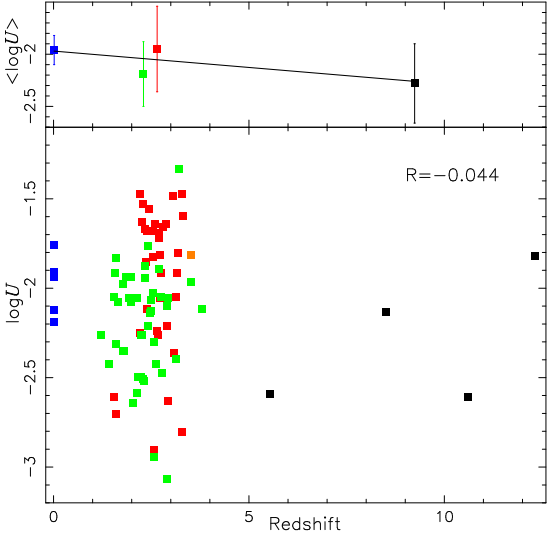}
 \caption{As Fig.~\ref{figcos} but for logarithm of the ionization parameter ($\log U$)
 estimated using the Eq.~\ref{eqc3c4}.}
\label{figcou}
\end{figure}

Firstly, let us analyze the dependence of $U$ with redshift. \citet{2013ApJ...774L..10K}
compared a large sample of SF spectroscopic data ($0.5 \: < \: z \: < \: 2.6$)
with predictions from photoionization models in a [\ion{O}{iii}]/H$\beta$ versus
[\ion{N}{ii}]/H$\alpha$ diagnostic diagram. These authors found that the ISM conditions are more extreme at high redshift than in local SFs; i.e., higher $U$ values are expected in high-$z$ SFs compared to local ones (see also \citealt{2018MNRAS.477.5568K, 2023ApJ...955...54S}).
However, in relation to AGNs, recent observations from the JADES spectroscopic survey carried out with the JWST by \citet{2025A&A...697A.175S} showed that the 42 identified type~2 AGNs
at $z \sim 10$ have a gas ionization degree similar to that found for objects at the local universe
(see also \citealt{2024MNRAS.531..355U, 2024arXiv241018193S, 2024arXiv240611997K, 2025arXiv250506359T, 2025arXiv250401852R}). To study this behavior, in Fig.~\ref{figcou}, bottom panel, results of
$\log U$ versus redshift for our type~2 AGN sample (see Table~\ref{tab1}) and for
the auxiliary objects (see Table~\ref{tab2}) are shown, separated by AGN classes. The $\log U$ values were derived using the
([\ion{C}{iii}]/\ion{C}{iv})-$U$ calibration 
represented by Eq.~\ref{eqc3c4}. We can see a large scattering in $U$ for fixed $z$ values,  
with a Pearson correlation coefficient $\rm R = -0.044$, i.e. there is no correlation between the estimates. 
In Fig.~\ref{figcou}, upper panel,
the mean $<\log U>$ parameter for different AGN classes is plotted against $z$.
A trend of slight $<\log U>$ decreases with increasing $z$ is observed. A linear fit to all $<\log U>$ values results in the following relation: 
\begin{equation}
    \log U= -0.03(\pm0.01) \times z -1.97(\pm 0.09).
\end{equation}
\textit{We emphasize that a large number of AGN estimates in distinct $z$ ranges (e.g. $z\sim1$ and $z\sim4$, 'redshift desert')
is necessary to confirm the $U$–$z$ relation derived above.} 

\begin{figure}
\includegraphics[angle=-90, width=0.4\textwidth]{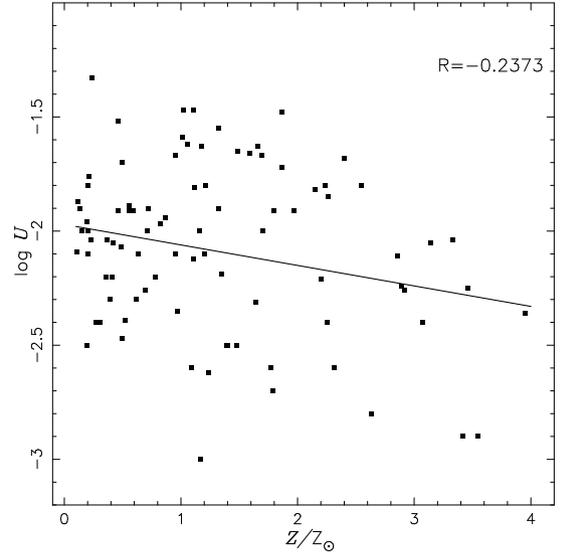}
 \caption{Logarithm of the ionization parameter $U$ versus the metallicity for our AGN sample (see Table~\ref{tab1}) derived using the Eqs.~\ref{eqc3c4} and \ref{eqc43a}, respectively. The line represents a fitting to the points (Eq.~\ref{equz}).}
\label{figcosn}
\end{figure}

Secondly, we analyze the relation between $\log U$ and $Z$, for which several
studies have shown contradictory results for SFs (e.g., \citealt{2011MNRAS.415.3616D, 2015A&A...574A..47S, 2016ApJ...822...42O, 2019MNRAS.486.1053K, 2022ApJ...937...22P, 2022A&A...659A.112J, 2022MNRAS.512.3436E, 2023ApJ...952..167R, 2023MNRAS.525.2087B, 2024A&A...684A..40P, 2025ApJ...978...70G, 2025A&A...695A..31L}) and AGNs (e.g., \citealt{2019MNRAS.489.2652P, 2024PASA...41...99O, 2024MNRAS.527.7217V}),
mainly due to the different methods (see \citealt{2008ApJ...681.1183K, 2020MNRAS.492..468D}) adopted to derive these parameters or the used samples.
In Fig.~\ref{figcosn}, a plot of $\log U$ versus $Z/\rm Z_{\odot}$, our
estimates for both object samples (see Tables~\ref{tab1} and \ref{tab2}), derived from the calibrations given by Eqs.~\ref{eqc3c4} and \ref{eqc43a}, respectively, are shown.
Objects are not separated by classes. We can note that $\log U$ tends to decrease with $Z$, following the
relation: 
\begin{equation}
\label{equz}
    \log U= -0.08(\pm0.03) \times (Z/\rm Z_{\odot}) -1.97(\pm0.06).
\end{equation}
This result is in agreement with the one obtained by \citet{2019MNRAS.489.2652P}, who using the \textsc{HCm} code, found
for local Seyfert~2
that $\log U$ tend to decrease with increasing $Z$.

\begin{figure}
\includegraphics[angle=-90, width=0.4\textwidth]{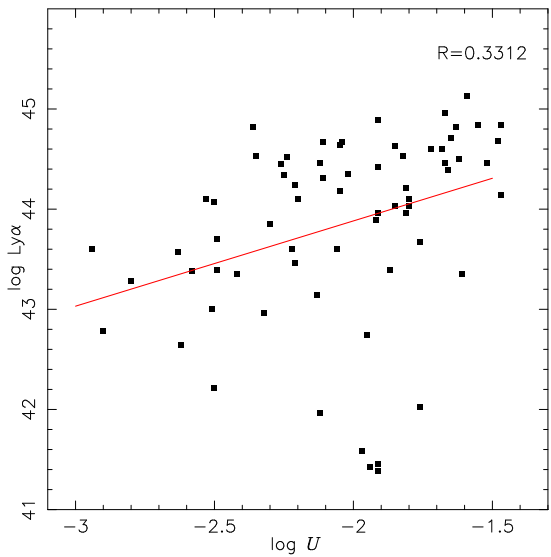}
 \caption{Logarithm of the ionization parameter versus the logarithm of the luminosity of  Ly$\alpha$ for our sample of AGNs (see Table~\ref{tab1}).  
 Values for $\log U$ were derived from Eq.~\ref{eqc3c4}.
 Pearson coefficient correlations (R) is presented in the plot. Red line represents a fitting to the points given by Eq.~\ref{equlya}.}
\label{fiana3f}
\end{figure}

Finally, in Fig.~\ref{fiana3f}, we analyze the dependence between the ionization parameter ($U$) and the Ly$\alpha$ luminosity for our sample. Again, the ionization parameter values were derived from the (\ion{C}{iii}/\ion{C}{iv})-$\log U$ calibration (Eq.\ref{eqc3c4}), using the observational data listed in Table~\ref{tab1}.
Despite a large scatter, a (weak) correlation (R = 0.3312) can be identified between the quantities, represented by
\begin{equation}
\label{equlya}
\log({\rm Ly\alpha}) [{\rm erg \: s^{-1}}]= 0.85 (\pm0.29)\times \log U + 45.58 (\pm0.61).
\end{equation}
From this analysis, we can assert that:
\begin{itemize}
\item Most of the AGNs in our sample could be classified as matter-bounded, because an increase in the number of ionizing photons ($U \sim Q(\rm H)$) leads to a direct increase in the 
Ly$\alpha$ luminosity, or in other words, in the ionized radius of the objects.
\item Objects with a high ionization parameter ($\log U \: \gtrsim \: -2.0$) and low luminosity ($\log \rm Ly\alpha \: \lesssim \: 42$ erg s$^{-1}$) could have a strong escape of ionizing radiation due to the combined effect of high $Q(\rm H)$ values and low gas content (radiation-bounded).
\end{itemize}
Both assertions would be confirmed in a future study based on detailed photoionization models assuming distinct AGN geometries and gas contents.

\subsection{Radio data correlations}

\citet{1969ApJ...157....1K} presented follow up observations of the \textit{Revised Third Cambridge Catalogue} \citep{1962MmRAS..68..163B}
and used the flux density ($S$) at two different frequencies 
($\nu_{1}$, $\nu_{2}$) as a definition of the spectral index:
\begin{equation}
\alpha_{\nu{1}}^{\nu_{2}} = \frac{\log(S_1/S_2)}{\log(\nu_1/\nu_2)}.
\end{equation}
Afterward, for example, \citet{1992ARA&A..30..575C} showed that
distinct object classes (e.g., active and non-active galaxies) present different $\alpha$ values (see also, 
\citealt{1997A&A...322...19N, 2000A&A...354..423L, 2017ApJ...836..185T, 2018MNRAS.474.5008D}, among others).

\begin{figure}
\includegraphics[angle=-90, width=0.4\textwidth]{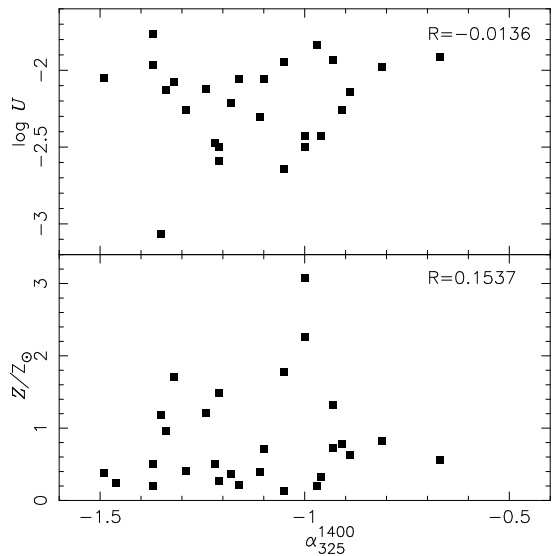}
 \caption{Metallicity (bottom panel) and logarithm of $U$ (upper panel) versus the radio spectral $\alpha_{325}^{1400}$ index (Eq.~\ref{eqsp}) for our AGN sample (see Table~\ref{tab1}). $Z$ and $U$ were derived from Eq.~\ref{eqc43a} and \ref{eqc3c4}, respectively. Pearson coefficient correlations (R) are indicated in each panel.}
\label{figra}
\end{figure}

The correlation between radio indexes and emission line
luminosities observed in most of radio galaxies and quasars
indicates that the central AGN is the common energy source for
both (e.g \citealt{1990ApJ...365..487M, 1999MNRAS.309.1017W})
as well as it shows a strong connection
between radio emission and conditions in the NLRs
(e.g. \citealt{2021ApJS..253...25K}).
However, \citet{2000A&A...362..519D} found that UV line ratios are not correlated with radio size
or radio power of High-$z$ radio galaxies, with the exception of the 
\ion{C}{ii}]$\lambda$2326/\ion{C}{iii}]$\lambda$1909 ratio, for which there is a  correlation  only for small radio sources, i.e. $D \lesssim$ 150 kpc (see also \citealt{2002MNRAS.337.1381I}). 
Moreover, \citet{2019A&A...630A..83Z} found that the location of radio sources in the narrow
emission line diagnostic diagrams shifts with the
increasing importance of a radio-loudness   AGN away from galaxies dominated by radio emission powered by star formation. Thus, this result indicates that the NLR conditions (e.g., $U$, $Z$, $\alpha_{ox}$), which drive 
the AGN position in BPT diagrams (e.g., \citealt{2014MNRAS.437.2376R, 2016MNRAS.456.3354F}), can be related to radio density fluxes. 
Since our sample is composed mainly ($\sim 60$ per cent) of radio galaxies, it is useful to verify the existence of spectral index correlations with  AGN parameters. In this context, we use the 
\begin{equation}
\label{eqsp}    \alpha_{325}^{1400}=\frac{\log[P(325 \: {\rm MHz})/P(1400\: {\rm  MHz})]}{\rm \log(325\: \rm MHz/\: 1400 \: MHz)}
\end{equation}
spectral index (using the radio data listed in Table~\ref{tab1}) to analyze possible correlations with $Z$ and $U$. In Fig.~\ref{figra}, metallicity (bottom panel) and the logarithm of the ionization parameter (upper panel) versus the $\alpha_{325}^{1400}$ for our 
sample of High-$z$ radio galaxies (29 objects) and Seyfert~2 (1 object) are 
shown.  We observe that, in agreement with \citet{2000A&A...362..519D}, no
correlation was derived, as indicated by the "R" values. 
Finally, we tested the dependence (not shown) between $Z$, $U$ and the 
radio power $P_{1400}$ and did not find any correlation.

\subsection{Mass-metallicity relation}

 \begin{figure}
\includegraphics[angle=-90, width=0.45\textwidth]{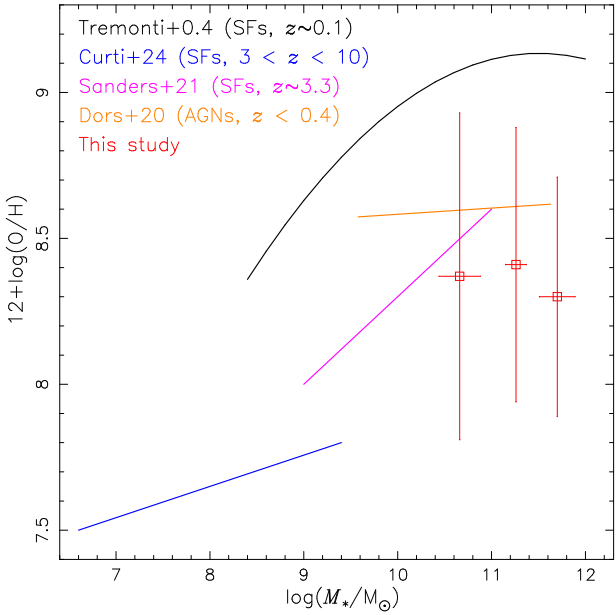}
\caption{Metallicity (traced by O/H abundance) of the NLR for our AGN sample (see Table~\ref{tab1}) versus the logarithm of stellar mass of the host
galaxy (in units of solar mass). Red  open squares represent mean metallicity for our sample derived considering estimates via
our calibration (Eq.~\ref{eqc43a}). Solid curves represent the relations for SF galaxies at distinct redshift and mass ranges by 
\citet{2004ApJ...613..898T}, \citet{2021ApJ...914...19S} and \citet{2024A&A...684A..75C} as indicated.
Also, the relation for local ($z \: < \: 0.4$) AGNs derived using observational data by 
\citet{2020MNRAS.492..468D}, whose $Z_{\rm AGN}$ was derived applying the empirical calibration by 
\citet{2021MNRAS.507..466D} is indicated.}
\label{figm}
\end{figure}

The pioneering study by \citet{1979A&A....80..155L} showed a clear and direct relation between the mass of galaxies and their metallicity (i.e mass-mettalicity relation, MZR). This relation arises due to more massive galaxies have deeper gravitational potential wells, which make it much harder for supernova- or AGN-driven winds to expel metals into the intergalactic medium. In contrast, low-mass galaxies lose a larger fraction of their metals through winds, so they tend to exhibit lower $Z$ compared to more massive systems (e.g., \citealt{2004ApJ...613..898T, 2018MNRAS.481.1690C, 2019MNRAS.487.3581F, 2021MNRAS.504...53S}).

Regarding the relation between the AGN metallicity ($Z_{\rm AGN}$) and the host galaxy mass, it could exist if the AGN chemical evolution is connected with that of the entire galaxy in which it is located. However, this relation is debated in the literature, and discrepant results have been derived, which can be due to the methods used to derive $Z$, the range of galaxy mass, and/or the redshift range of the sample. For instance, \citet{2018A&A...616L...4M}, who estimated $Z$ from UV emission lines for a sample of type~2 AGNs ($1.2 \: < \: z \: < \: 4.0$), showed that there is a direct relation between the NLR metallicities and the stellar masses ($M_{\star}$) of their host galaxies (see also \citealt{2019MNRAS.486.5853D}). However, \citet{2020MNRAS.492..468D}, who estimated $Z$ via optical emission lines for a large sample (463 objects) of Seyfert~2 nuclei in the local universe ($z \: < \: 0.4$), did not find any relation between $Z_{\rm AGN}$ and $M_{\star}$ of the host galaxy, independent of the method used to derive the metallicity (see also \citealt{2024PASA...41...99O, 2024MNRAS.529.4993L}).

 Since our new $C43$-$Z$ calibration provides reliable values, i.e., in agreement with those obtained via the $T_{\rm e}$ method, it is desirable to investigate the $Z_{\rm AGN}$-$M_{\star}$ relation for the sample of objects considered in the present study. In this context, we divide the sample into bins of 0.5 dex for the stellar mass 
 ($10.67 \: \lesssim  \: \log(\frac{M_{\star}}{\rm M_{\odot}}) \:   \lesssim\: 11.7$)  and calculate the mean value of $Z$. For a total of 18 objects ($1.5 \: < \: z \: < \: 3.6$) in our sample, it was possible to consider estimates for $Z_{\rm AGN}$ and $M_{\star}$. This reduced number is due to the lack of $M_{\star}$ combined with the valid range of $Z$ defined by our metallicity calibration (see Table \ref{tab1}). In Fig.~\ref{figm}, a plot of 12+log(O/H)\footnote{We convert metallicity into O/H abundance by the expression $12+\log({\rm O/H})= 12+\log[(Z/\rm Z_{\odot}) \times 10^{-3.31}]$,
 where $\rm \log(O/H)_{\odot}=-3.31$ \citep{2001ApJ...556L..63A}.}   
 versus $\log(M_{\star}/\rm M_{\odot})$, shows the results for our estimates (red open squares). For comparison, we also plot several mass--metallicity relations from the literature:
\begin{itemize}
    \item The mass--metallicity relation for $\sim 53\:000$ SF galaxies ($z\sim0.1$) derived by \citet{2004ApJ...613..898T} and valid for $8.5 \: < \: \log(M_{\star}/{\rm M_{\odot}}) \: < \: 11.5$. These authors used strong-line methods to estimate $Z$. 
    \item The mass--metallicity relation 
    derived by \citet{2024A&A...684A..75C} for
    SF galaxies at $3 \: < \: z \: < \: 10$
    and with $\log(M_{\star}/{\rm M_{\odot}}) \: \lesssim\: 9$. The $Z$ values were estimated through strong-line methods.
    \item The mass-metallicity relation derived by \citet{2021ApJ...914...19S} for SF galaxies at $z\sim3.3$ by using $Z$ estimates from strong-line methods. The mass interval considered by these authors is $9 \: < \: \log(M_{\star}/{\rm M_{\odot}}) \: < \: 10.5$.
    \item The $M_{\star}$-$Z_{\rm AGN}$ derived using  spectroscopic data by 
    \citet{2020MNRAS.492..468D} for $\sim400$ AGNs at $z\: < \: 0.4$. The $Z_{\rm AGN}$ for each object was derived through the $R_{23}$-(O/H) empirical calibration proposed by \citet{2021MNRAS.507..466D}.
    This derived relation is valid for  $9.5 \: < \: \log(M_{\star}/{\rm M_{\odot}}) \: < \: 11.5$.    
\end{itemize}
We can see in Fig.~\ref{figm} that our $Z_{\rm AGN}$-$M_{\star}$ estimations (red open squares) for AGNs at $1.5 \: < \: z \: < \: 3.6$ is flat, which does not indicate any dependence between the AGN metallicity and $M_{\star}$. Our results are in agreement with estimations for local ($z \: < \: 0.4)$ AGNs derived by \citet{2020MNRAS.492..468D}. It is worth noting that the AGNs of our sample are located in more massive galaxies than those considered by \citet{2024A&A...684A..75C} and \citet{2021ApJ...914...19S}, while the former studied galaxies in a wider redshift range than the objects in our sample and the last authors only considered objects at $z\sim 3.3$. In both cases, they derived steep relations between $Z$ and $M_{\star}$. Finally, comparing our estimations  with the relation derived by
\citet{2004ApJ...613..898T}, we can see that AGNs have a lower mean metallicity compared to SFs with similar masses.  From a galaxy evolution perspective, the flat $Z_{\rm AGN}$-$M_{\star}$ relation presented in Fig.~\ref{figm} is a significant result.  It contrasts sharply with the well-established positive MZR for star-forming galaxies, where the deeper gravitational potential wells of more massive galaxies are better able to retain the metals produced by supernovae.

We emphasize that in some previous studies,  in which
the AGN metallicity is derived from UV emission lines (e.g. \citealt{2018A&A...616L...4M, 2019MNRAS.486.5853D}),
 an increase of $Z_{\rm AGN}$ with $M_{\star}$ has been derived. However, most of those $Z$ estimates were performed using carbon UV emission line intensities and photoionization model predictions assuming a C/O-O/H abundance relation that may not be representative of AGNs.
 In contrast, in the present study the C/O-O/H relation derived from AGNs and SFs based on the $T_{\rm e}$-method is assumed in the photoionization models.  The sample used in Paper I to constrain the relation for active nuclei included only 7 AGNs. Although this number is small, it represents the first such relation for AGNs derived using the most reliable direct method. Our current work is an application of this new, physically motivated relation to test its implications on a larger sample of objects. We emphasize that refining this C/O–O/H relation with a larger sample of AGNs with direct-method abundances is a critical next step for the field. Nevertheless, the C/O–O/H relation assumed in our models is expected to provide, in principle, more reliable results than those obtained using previous methods (see \citealt{2025MNRAS.540.1608D}).

\section{Conclusion}
\label{conc}

In a previous study, \citet{2025MNRAS.540.1608D} (Paper~I) used abundance estimates obtained  through the $T_{\rm e}$-method (the most reliable method) to derived a C/O-O/H relation representative for narrow line regions of AGNs. Assuming this abundance relation in a grid of photoionization models
and using ultraviolet spectroscopic data (compiled from the literature) of 106 type~2 AGNs
($0 \: \lesssim \: z \: \lesssim  \:  4$), we obtained a new semi-empirical calibration between the $C43$=(\ion{C}{iv}$\lambda1549$+\ion{C}{iii}]$\lambda1909$)/\ion{He}{ii}$\lambda1640$ line ratio and $Z$ for the NLRs of AGNs. A calibration between the  $C3C4$=(\ion{C}{iii}]$\lambda1909$/\ion{C}{iv}$\lambda1549$) line ratio and the ionization parameter $U$ was also proposed.
Applying these new calibrations we obtain the following results and conclusions:

\begin{itemize}
    \item  Metallicity derived through the new calibration is consistent with those derived
    through the $T_{\rm e}$-method and optical strong-line methods. Although the comparison of $Z$ derived by  methods based on optical emission lines is possible for few objects, the consistence between the $Z$ values indicate that the new $C43$-$Z$ calibration results in reliable values.
\item Combining  auxiliary observational data of AGNs at $5 \: \lesssim \: z \: \lesssim \: 12$ with those of
    our sample, we find no evidence for a monotonic evolution of AGN metallicity across the redshift range 
$0 \: \lesssim \: z \: \lesssim \: 12$. This result indicates that AGNs had the main star formation episode  and chemical enrichment  of the ISM in an early epoch $(z \: > \: 5)$
of their formation.  The most striking feature is the evidence for very high metallicities at all epochs. Notably, the highest metallicities in our sample, reaching up to $4\,Z_{\odot}$, are found in luminous Type~2 Quasars and high-redshift radio galaxies at $2\: \lesssim \: z\: \lesssim \:3$. As shown in Fig.~\ref{figcos}, this epoch coincides with the peak of the cosmic star formation rate history ($\psi$). This finding strongly suggests a powerful link between the `cosmic noon' of star formation, rapid supermassive black hole growth, and an intense, early chemical enrichment phase in the host galaxies of luminous AGNs.
   \item We found a slight decrease of the ionization parameter $U \approx Q({\rm H})/N_{\rm e}$
   with the increase of the redshift. This result indicates that either the number of ionization photons decrease with $z$ or AGNs at higher redshifts present higher $N_{\rm e}$ values compared to
   those in the local universe ($z \: < \: 0.4$).
   \item  We found no significant correlations between radio data [spectral index $\alpha_{325}^{1400}$ and luminosity $L(1400$ MHz)] and
   $Z$, as well as $U$ of the AGNs, which indicates that the central AGN is not the common energy source for ionized and neutral gas regions.  
   \item 
   We find no statistically significant evidence for a positive MZR, and that the data are consistent with a flat relation.  The lack of a mass-metallicity dependence for AGNs, both at high redshift (this study) and in the local universe, suggests that the chemical evolution of the gas in the immediate vicinity of the supermassive black hole is decoupled from the global chemical evolution of the host galaxy. We note here that a larger and more homogeneous sample is required to confirm this result.
\end{itemize}
 
\section*{Acknowledgements}
OLD is grateful to Funda\c cão de Amparo à
Pesquisa do Estado de São Paulo (FAPESP) and Conselho Nacional
de Desenvolvimento Científico e Tecnológico (CNPq). MA gratefully acknowledges support from Fundação de Amparo à Pesquisa do Estado de São Paulo (FAPESP, Processo:	
2024/03727-3).
\section*{Data Availability}

The data underlying this article will be shared on reasonable request
to the corresponding author.

\bibliographystyle{mnras}
\bibliography{refs} 







\bsp	
\label{lastpage}
\end{document}